\documentclass[sigconf]{acmart}
\usepackage[ruled,noend]{algorithm2e}
\usepackage{listings}
\usepackage{xcolor}
\usepackage{multirow}
\usepackage{enumitem}
\usepackage[flushleft]{threeparttable}
\usepackage{graphicx}
\usepackage{subfigure}
\usepackage{cleveref}
\usepackage{mathtools}
\usepackage{nicefrac}

\acmConference[Under review of SIGKDD]{}{2022}{Washington DC}

\AtBeginDocument{%
  \providecommand\BibTeX{{%
    \normalfont B\kern-0.5em{\scshape i\kern-0.25em b}\kern-0.8em\TeX}}}

\setcopyright{none}
\renewcommand\footnotetextcopyrightpermission[1]{}
\begin{document}

\title{Approximate Nearest Neighbor Search under Neural Similarity Metric for Large-Scale Recommendation}

\author{Rihan Chen, Bin Liu, Han Zhu, Yaoxuan Wang, Qi Li, Buting Ma}
\author{Qingbo Hua, Jun Jiang, Yunlong Xu, Hongbo Deng, Bo Zheng}
\email{{rihan.crh, zhuoli.lb, zhuhan.zh, wangyaoxuan.wyx, luyuan.lq, buting.mbt}@alibaba-inc.com}
\email{{yuxing.hqb, chuncao.jj, yunlong.xyl, dhb167148, bozheng}@alibaba-inc.com}
\affiliation{%
  \institution{Alibaba group}
  \state{Beijing}
  \country{China}
}

\renewcommand{\shortauthors}{Rihan and Bin, et al.}
\newcommand{\argmax}{\operatornamewithlimits{argmax}}
\newcommand{\argmin}{\operatornamewithlimits{argmin}}
\newcommand{\argTopk}{\operatornamewithlimits{argTopk}}
\newcommand{\argTopm}{\operatornamewithlimits{argTopm}}

\newcommand{\epsilonv}{\boldsymbol \epsilon}
\newcommand{\thetav}{\boldsymbol \theta}
\newcommand{\Thetav}{\boldsymbol \Theta}
\newcommand{\phiv}{\boldsymbol \phi}
\newcommand{\Phiv}{\boldsymbol \Phi}
\newcommand{\betav}{\boldsymbol \beta}
\newcommand{\Betav}{\boldsymbol \Beta}
\newcommand{\etav}{\boldsymbol \eta}
\newcommand{\Etav}{\boldsymbol \Eta}
\newcommand{\lambdav}{\boldsymbol \lambda}
\newcommand{\Lambdav}{\boldsymbol \Lambda}
\newcommand{\omegav}{\boldsymbol \omega}
\newcommand{\Omegav}{\boldsymbol \Omega}
\newcommand{\alphav}{\boldsymbol \alpha}
\newcommand{\Alphav}{\boldsymbol \Alpha}
\newcommand{\deltav}{\boldsymbol \delta}
\newcommand{\Deltav}{\boldsymbol \Delta}
\newcommand{\sigmav}{\boldsymbol \sigma}
\newcommand{\Sigmav}{\boldsymbol \Sigma}
\newcommand{\muv}{\boldsymbol \mu}
\newcommand{\Muv}{\boldsymbol \Mu}
\newcommand{\nuv}{\boldsymbol \nu}
\newcommand{\Nuv}{\boldsymbol \Nu}
\newcommand{\zerov}{\boldsymbol 0}
\newcommand{\onev}{\boldsymbol 1}

\newcommand{\av}{\mathbf{a}}
\newcommand{\bv}{\mathbf{b}}
\newcommand{\cv}{\mathbf{c}}
\newcommand{\dv}{\mathbf{d}}
\newcommand{\ev}{\mathbf{e}}
\newcommand{\fv}{\mathbf{f}}
\newcommand{\gv}{\mathbf{g}}
\newcommand{\hv}{\mathbf{h}}
\newcommand{\iv}{\mathbf{i}}
\newcommand{\jv}{\mathbf{j}}
\newcommand{\kv}{\mathbf{k}}
\newcommand{\lv}{\mathbf{l}}
\newcommand{\mv}{\mathbf{m}}
\newcommand{\nv}{\mathbf{n}}
\newcommand{\ov}{\mathbf{o}}
\newcommand{\pv}{\mathbf{p}}
\newcommand{\qv}{\mathbf{q}}
\newcommand{\rv}{\mathbf{r}}
\newcommand{\sv}{\mathbf{s}}
\newcommand{\tv}{\mathbf{t}}
\newcommand{\uv}{\mathbf{u}}
\newcommand{\wv}{\mathbf{w}}
\newcommand{\xv}{\mathbf{x}}
\newcommand{\yv}{\mathbf{y}}
\newcommand{\zv}{\mathbf{z}}
\newcommand{\Av}{\mathbf{A}}
\newcommand{\Bv}{\mathbf{B}}
\newcommand{\Cv}{\mathbf{C}}
\newcommand{\Dv}{\mathbf{D}}
\newcommand{\Ev}{\mathbf{E}}
\newcommand{\Fv}{\mathbf{F}}
\newcommand{\Gv}{\mathbf{G}}
\newcommand{\Hv}{\mathbf{H}}
\newcommand{\Iv}{\mathbf{I}}
\newcommand{\Jv}{\mathbf{J}}
\newcommand{\Kv}{\mathbf{K}}
\newcommand{\Lv}{\mathbf{L}}
\newcommand{\Mv}{\mathbf{M}}
\newcommand{\Nv}{\mathbf{N}}
\newcommand{\Ov}{\mathbf{O}}
\newcommand{\Pv}{\mathbf{P}}
\newcommand{\Qv}{\mathbf{Q}}
\newcommand{\Rv}{\mathbf{R}}
\newcommand{\Sv}{\mathbf{S}}
\newcommand{\Tv}{\mathbf{T}}
\newcommand{\Uv}{\mathbf{U}}
\newcommand{\Vv}{\mathbf{V}}
\newcommand{\Wv}{\mathbf{W}}
\newcommand{\Xv}{\mathbf{X}}
\newcommand{\Yv}{\mathbf{Y}}
\newcommand{\Zv}{\mathbf{Z}}
\DeclarePairedDelimiter\ceil{\lceil}{\rceil}
\DeclarePairedDelimiter\floor{\lfloor}{\rfloor}
\newcommand{\zhuoli}[1]{\textcolor[rgb]{0.545,0.545,0}{(Zhuoli: #1)}}

\definecolor{mygreen}{rgb}{0,0.6,0}
\definecolor{mygray}{rgb}{0.5,0.5,0.5}
\definecolor{mymauve}{rgb}{0.58,0,0.82}

\lstset{ %
  backgroundcolor=\color{white},   %
  basicstyle=\footnotesize,        %
  breaklines=true,                 %
  captionpos=b,                    %
  commentstyle=\color{mygreen},    %
  escapeinside={\%*}{*)},          %
  keywordstyle=\color{blue},       %
  stringstyle=\color{mymauve},     %
}

\begin{abstract}
Model-based methods for recommender systems have been studied extensively for years. Modern recommender systems usually resort to 1) representation learning models which define user-item preference as the distance between their embedding representations, and 2) embedding-based Approximate Nearest Neighbor (ANN) search to tackle the efficiency problem introduced by large-scale corpus. While providing efficient retrieval, the embedding-based retrieval pattern also limits the model capacity since the form of user-item preference measure is restricted to the distance between their embedding representations. However, for other more precise user-item preference measures, e.g., preference scores directly derived from a deep neural network, they are computationally intractable because of the lack of an efficient retrieval method, and an exhaustive search for all user-item pairs is impractical. 

In this paper, we propose a novel method to extend ANN search to arbitrary matching functions, e.g., a deep neural network. Our main idea is to perform a greedy walk with a matching function in a similarity graph constructed from all items. To solve the problem that the similarity measures of graph construction and user-item matching function are heterogeneous, we propose a pluggable adversarial training task to ensure the graph search with arbitrary matching function can achieve fairly high precision. Experimental results in both open source and industry datasets demonstrate the effectiveness of our method. The proposed method has been fully deployed in the Taobao display advertising platform and brings a considerable advertising revenue increase. We also summarize our detailed experiences in deployment in this paper.

\end{abstract}

\ccsdesc[500]{Information systems~Recommender systems}
\ccsdesc[500]{Information systems~Personalization}

\keywords{Approximate Nearest Neighbor Search, Model-based Retrieval,  Recommender Systems}

\maketitle

\section{Introduction}
Constantly growing amount of available information has posed great challenges to modern recommenders. To deal with the information explosion, modern recommender system is usually designed with a multi-stage cascade architecture that mainly consists of candidate generation and ranking. In the candidate generation stage, also known as matching, thousands of targets are retrieved from a very large corpus, and then, in the ranking stage, these retrieved targets are ranked according to the user's preference. Notably, given the constraints of computational resources and latency in real-world systems, candidate generation cannot be solved by sequentially scanning the entire corpus while facing a large-scale corpus.

To bypass the prohibitive computational cost of scanning the entire corpus, embedding-based retrieval (EBR) has prevailed in recommender systems for years due to its simplicity and efficiency~\cite{huang2020embedding, li2019multi}. 
However, EBR is insufficient to model the complex structure of user-item preferences. Many works have already shown that more complex models usually generalize better~\cite{he2017neural, zhou2018deep, pi2020search}. And researchers have striven to develop techniques to tackle the large-scale retrieval problem with more complex models as well.
To overcome computation barriers and benefit from arbitrarily advanced models, the idea of regularizing the total computational cost through an index has recently been presented. These methods~\cite{zhu2018learning, zhu2019joint, zhuo2020learning, gao2020deep} typically have a learnable index and follow the Expectation Maximization (EM) type optimization paradigm, updating between deep model and index alternatively. 
As a consequence, the deep model, together with beam search, can be leveraged to retrieve relevant items from a large corpus in a sub-linear complexity w.r.t. corpus size. Even though these end-to-end methods can introduce a deep model to large-scale retrieval, there are two aspects that should not be ignored: 1) the joint training of index and model for large-scale data necessitates a costly training budget in terms of both training time and computational resources; 2) 
the existence of index structure's internal nodes, such as non-leaf nodes in TDMs~\cite{zhu2018learning, zhu2019joint, zhuo2020learning} and path nodes in DR~\cite{gao2020deep}, makes it difficult to utilize side-information from items.

This work tackles the aforementioned problems by solving large-scale retrieval with an arbitrarily advanced model in a lightweight manner, called Neural Approximate Nearest Neighbour Search (NANN). More specifically, we leverage the deep model as a greedy walker to explore a similarity graph constructed after model training.
The joint training budget of the end-to-end methods can be greatly released by following the decoupled paradigm.
Besides, the similarity graph that the deep model traverses contains no internal nodes, which facilitates the usage of side information from candidate items. To improve the efficiency and effectiveness of graph search, we creatively come up with both a heuristic retrieval method and a auxiliary training task in our NANN framework. 
The main contributions of our paper are summarized as follows:
\begin{itemize}
    \item We present a unified and lightweight framework that can introduce arbitrarily advanced models as the matching function to large-scale ANN retrieval. The basic idea is to leverage similarity graph search with the matching function.
    \item To make the computational cost and latency controllable in graph search, we propose a heuristic retrieval method called Beam-retrieval, which can reach better results with fewer computations. And we also propose an auxiliary adversarial task in model training, which can greatly mitigate the effect of heterogeneity between similarity measures and improve the retrieval quality.
    \item We conduct extensive experiments on both a publicly accessible benchmark dataset and a real industry dataset, which demonstrate the proposed NANN is an excellent empirical solution to ANN search under neural similarity metric. Besides, NANN has been fully deployed in the Taobao display advertising platform and contributes 3.1\% advertising revenue improvements.
    \item We describe in detail the hands-on deployment experiences of NANN in Taobao display advertising platform. The deployment and its corresponding optimizations are based on the Tensorflow framework~\cite{abadi2016tensorflow}. We hope that our experiences in developing such a lightweight yet effective large-scale retrieval framework will be helpful to outstretch NANN to other scenarios with ease.
\end{itemize}

\section{Related Work}
Hereafter, let $\mathcal{V}$ and $\mathcal{U}$ denote the item set and the user set. In recommendation, we strive to retrieve a set of relevant items $\mathcal{B}_u$ from a large-scale corpus $\mathcal{V}$ for each user $u \in \mathcal{U}$. Mathematically,
\begin{equation}\label{eq:topk}
\mathcal{B}_u = \argTopk_{v \in \mathcal{V}} s(v, u)
\end{equation}
where $s(v, u)$ is the similarity function.

\textbf{Search on graph}. Search on graph popularized by its exceptional efficiency, performance, and flexibility (in terms of similarity function) is a fundamental and powerful approach for NNS. The theoretical foundation to search on graph is the $s$-Delaunay graph defined by similarity function $s(v, u)$. Previous work~\cite{navarro2002searching} has shown that $s(v, u) = -||v-u||_2$ can find the exact solution to~\Cref{eq:topk}, when $k=1$ and $u, v \in \mathbb{R}^d$, by certain greedy walk on the $s$-Delaunay graph constructed from $\mathcal{V}$. More generally, many existing works attempt to extend the conclusion to non-metric cases, such as inner product~\cite{bachrach2014speeding, ram2012maximum, shrivastava2014asymmetric, shrivastava2015asymmetric}, Mercer kernel~\cite{curtin2013fast, curtin2014dual} and Bregman divergence~\cite{cayton2008fast}. In addition, researchers also set foot in approximating the $s$-Delaunay graph as the construction of a perfect $s$-Delaunay graph with a large corpus is infeasible. Navigable Small World (NSW)~\cite{malkov2014approximate} is proposed to greatly optimize both graph construction and search process. On top of that, Hierarchical NSW (HNSW)~\cite{malkov2018efficient} incrementally builds a multi-layer structure from proximity graphs and provides state-of-the-art for NNS. Our approach will resort to HNSW, although other graph-based NNS methods can also work.

\textbf{Deep model-based retrieval}. Model-based, especially deep model-based methods have been an active topic in large-scale retrieval recently. In recommendation, many works focus on an end-to-end fashion to simultaneously train index and deep model. Tree-based methods, including TDM~\cite{zhu2018learning}, JTM~\cite{zhu2019joint} and BSAT~\cite{zhuo2020learning}, build its index as a tree structure and model user interests from coarse to fine. Deep retrieval (DR)~\cite{gao2020deep} encodes all candidate items with learnable paths and train the item paths along with the deep model to maximize the same objective. These approaches traverse their index to predict user interests and achieve sub-linear computational complexity w.r.t corpus size by beam search. However, these methods usually require additional internal nodes to parametrize the learnable index, which imposes difficulties in using side information of items. Moreover, additional model parameters and training time have to be paid for these end-to-end manners due to the existence of a learnable index and EM-type training paradigm. 

\textbf{Search on the graph with deep model} A few works have already tried to extend the similarity function to deep neural networks. The closest work to ours is SL2G~\cite{tan2020fast} which constructs the index graph by l2 distance and traverses the post-training graph with deep neural networks. However, their approach can be only generalized to the $s(v, u)$ with convexity or quasi-convexity. For the non-convex similarity function (most common case for deep neural network), they apply SL2G directly without adaption. %
Another work \cite{morozov2019relevance} defines the index graph without similarity for item pairs. They exploit the idea that relevant items should have close $s(v, u)$ for the same user and represent a candidate item by a subsample of $\{s(v, u_j)|j=1,\dots,m\}$. However, it is difficult to sample a representative set in practice, especially for large-scale corpus $\mathcal{V}$.

\section{Methodology}
In this section, we firstly give a general framework about EBR and model-based retrieval in~\Cref{sec:general_framework}, including model architecture and training paradigm. Then, we introduce the similarity graph construction and graph-based retrieval method respectively for the proposed NANN in~\Cref{sec:graph_construction} and~\Cref{sec:online_retrieval}. Given these preliminary concepts, we accordingly introduce the pluggable adversarial training task and demonstrate how it can eliminate the gap of similarity measures between graph construction and model-based matching function in \Cref{sec:neural_metric}.
\begin{figure*}
    \centering
    \includegraphics[width=0.9\textwidth]{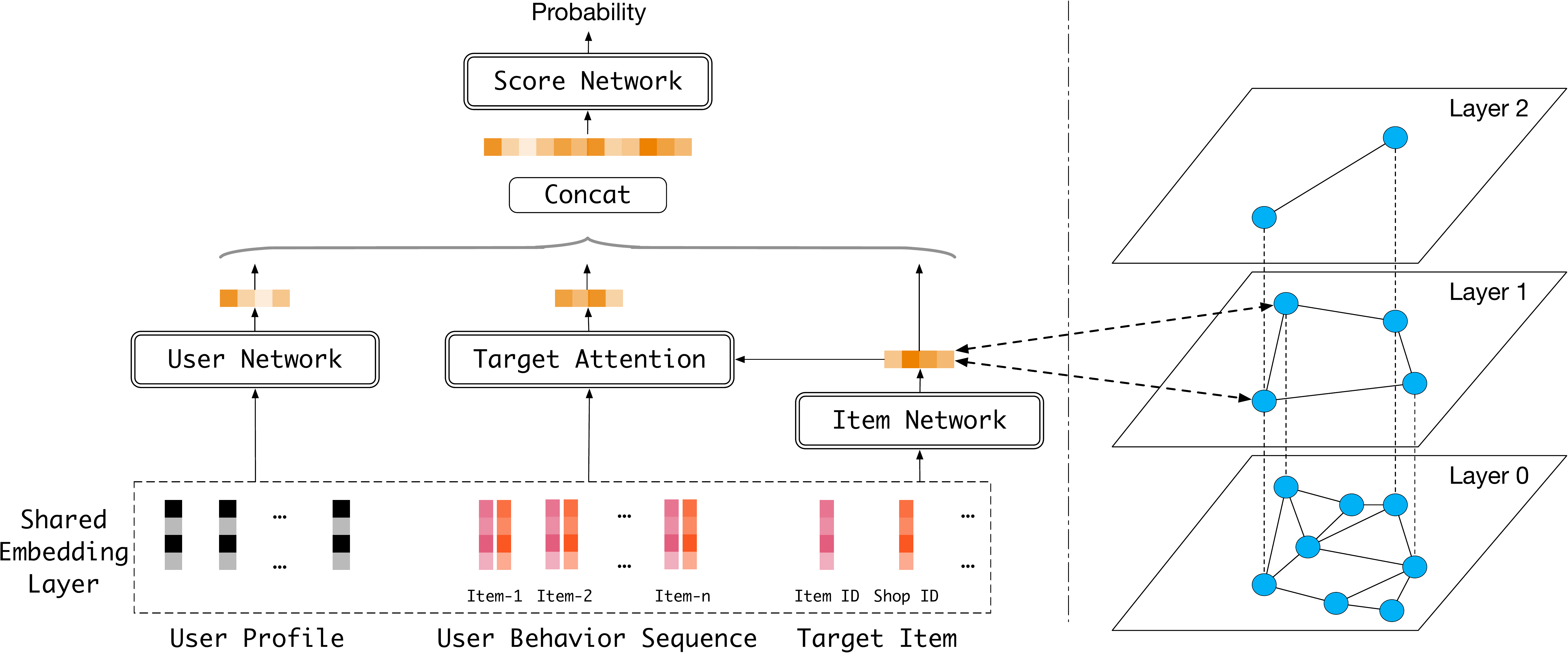}
    \vspace{-5pt}
    \caption{General Framework. In the left part, the deep model contains three basic branches, user network, target attention network, and item network. The user network is responsible for modeling the user's profile. And we adopt the attention mechanism to flexibly learn the interaction between user behavior sequence and target item. The information of items is learned through the item network. In the right part, we approximate the Delaunay graph defined on $l2$ distance among $\ev_v$ (the output of item network) following HNSW. Note that the online inference starts from $\ev_v$ for item network branch.}
    \label{fig:network}
    \vspace{-10pt}
\end{figure*}

\subsection{General Framework}\label{sec:general_framework}
\subsubsection{Review Embedding Based Retrieval} Our proposed method can be generally deemed as an extension of the EBR framework where we generalize the simple similarity metrics to arbitrary neural ones. Therefore, we briefly review the EBR framework for clarity.

EBR is designed with a two-sided model architecture where one side is to encode the user profile and behaviour sequence, and the other side is to encode the item.  Mathematically, 
\begin{equation}\label{eq:ebr}
    \ev_u = \mathrm{NN}_u(\fv_u),~~\ev_v = \mathrm{NN}_v(\fv_v).
\end{equation}
where two deep neural networks $\mathrm{NN}_u$ and $\mathrm{NN}_v$ (i.e., the user and the item network) encode the inputs of $\fv_u$ and $\fv_v$ to the dense vectors  $\ev_u \in \mathbb{R}^d$ and $\ev_v \in \mathbb{R}^d$ separately. And the user-item preference forms as the inner product of the semantic embedding, i.e. $\ev_u^T\ev_v$. The candidate sampling  based criterion such as Noise Contrastive Estimation (NCE)~\cite{gutmann2010noise} and Sampled-softmax~\cite{jang2016categorical} are usually used to train the EBR models due to the computational difficulty to evaluate partition functions by summing over the entire vocabulary of large corpus.

\subsubsection{Model Architecture}
Compared to the traditional EBR method in large-scale retrieval, NANN greatly outstretches the model capacity by more complex architecture with user network, target attention network, and item network akin to a standard CTR prediction model, as shown in \Cref{fig:network}. 
 In other words, we substitute the inner product $\ev_u^T\ev_v$ with a more general and expressive $s(v, u)$. The generalized form $s(v, u)$ with deep neural network, in turn, poses both theoretical and practical challenges to us: 1) how to generalize the search on the graph-based index to any non-linear and non-convex $s(v, u)$ reasonably; 2) how to integrate the graph-based index with complex deep model and deploy the computation-intensive retrieval framework in a lightweight and efficient way. 

\subsubsection{Training}
Same with EBR, we reduce the computationally intractable problem to the problem of estimating the parameters of a binary classifier by NCE. The positive samples come from the true distribution that user $u$ engages with item $v$, while the negative samples are drawn from a “noise” distribution $q(v)$, e.g., the unigram distribution over $v\in\mathcal{V}$. We denote the corresponding loss function as $ \mathcal{L}_{NCE}$. Moreover, we extend the search on the graph-based index to any metric $s(v, u)$ by using an auxiliary task with the loss denoted by $\mathcal{L}_{AUX}$ (details are in~\Cref{sec:neural_metric}). Hence, the overall objective is
\begin{equation}\label{eq:loss_all}
\mathcal{L}_{all} =  \mathcal{L}_{NCE} + \mathcal{L}_{AUX}.
\end{equation}

\subsubsection{Search on post-training similarity graph} The graph-based index is built from the precomputed item embedding $\ev_v$ extracted from item network $\mathrm{NN}_v$. In the prediction stage, we traverse the similarity graph in a way that is tailored to both real-world systems and arbitrary $s(v, u)$.
 
\subsection{Graph Construction}\label{sec:graph_construction}
Search on similarity graphs was originally proposed for metric spaces and extended to the symmetric non-metric scenarios, e.g, Mercer kernel and Maximum Inner Product Search (MIPS). The $s(v, u)$ can be also generalized to the certain asymmetric case, i.e. Bregman divergence, by exploiting convexity in place of triangle inequality~\cite{cayton2008fast}. However, $s$-Delaunay graph with arbitrary $s(v, u)$ is not guaranteed to exist or be unique. Furthermore, to construct such $s$-Delaunay graphs from the large-scale corpus are even computationally prohibitive for both exact and approximate ones. Hence, we follow the way of SL2G \cite{tan2020fast} to simplify this problem by building the graph index with the item embedding $\ev_v$. The graph is defined with $l2$ distance among $\ev_v$ and agnostic to $s(v, u)$. In practice, we build the HNSW graph directly, which is claimed a proper way to approximate the Delaunay graph defined on $l2$ distance.

\subsection{Online Retrieval}\label{sec:online_retrieval}
We equip the original HNSW with beam search and propose a Beam-retrieval to handle the online retrieval in production.

With precomputed $\ev_v$, the online retrieval stage can be represented as
\begin{equation}\label{eq:online_retrieval}
\mathcal{B}_u = \argTopk_{v \in \mathcal{V}} s_u(\ev_{v}), 
\end{equation}
where $s_u(.)$ is the user specfic function computed in real time and $\ev_{v}$ is the only variable w.r.t $s_u(.)$ when search on graph-based index.

The search process of HNSW traverses a hierarchy of proximity graphs in a layer-wise and top-down way, as shown in~\Cref{algorithm:total}. The original HNSW retrieval algorithm referred to as HNSW-retrieval for convenience, employs simple greedy searches where $ef_l (l > 0)$ in~\Cref{algorithm:total} is set to 1 at the top layers and assigns a larger value to $ef_0$ to ensure retrieval performance at the ground layer. However, the HNSW-retrieval is practically insufficient to tackle large-scale retrieval in real-world recommender systems since it suffers from the following deficiencies: 1) the subroutine $\rm{SEARCH-LAYER}$ in HNSW-retrieval explores the graph in a while-loop, which makes the online inference's computation and latency uncontrollable; 2) the traversal with simple greedy search is more prone to stuck into local optimum, especially for our case where $s_u(\ev_v)$ is usually non-convex. Hence, we reform the $\rm{SEARCH-LAYER}$ subroutine in HNSW-retrieval according to~\Cref{algorithm:dvf}. We firstly replace the while-loop with a for-loop to control the prediction latency and the amount of candidates $v$ to evaluate. Despite having an early-stopping strategy, the for-loop can still guarantee the retrieval performance, shown in~\Cref{fig:rounds_results}. We secondly break the limits on $ef_l (l > 0)$ and enlarge it at top layers to utilize batch computing. Traversal with multiple paths is equivalent to beam search on the similarity graph, which is proved more efficient than the original version demonstrated in \Cref{fig:main_results}. 

\begin{algorithm}[!tp]
    \DontPrintSemicolon
    \caption{K-NN-SEARCH($s, G, u, K, ef$)}
    \KwIn{model $s$, multi-layer HNSW graph $G$, query user $u$, number of nearest neighbors to return $K$, size of the dynamic candidate list and number of steps to search for each layer $\{ ef_l \}, \{ T_l \}$}
    \KwOut{$K$ results with largest score $s(u, v)$ to user $u$}
    $W$ $\leftarrow$ $\emptyset$  \tcp{the set of the current nearest elements}
    $ep$ $\leftarrow$ get enter points from HNSW graph $G$\;
    $L \leftarrow$ level of $ep$ \tcp{the top layer of HNSW graph}
    \For{$l \leftarrow L \cdots 1$} {
    	$W \leftarrow$ SEARCH-LAYER($s, u, ep, ef_l, l=l, T_{l}$)\;
    	$ep \leftarrow W$\;
    }
	$W \leftarrow$ SEARCH-LAYER($s, u, ep, ef_0, l=0, T_0$)\;
	\KwRet{$\text{argtopk}_{v \in W} f(u, v)$}
	\label{algorithm:total}
\end{algorithm}

\begin{algorithm}[!tp]
    \DontPrintSemicolon
    \caption{SEARCH-LAYER($s, u, ep, ef_c, l_c, T_c$)}
    \KwIn{enter points $ep$, current layer number $l_c$, number of steps to search in this layer $T_c$}
    \KwOut{$ef_c$ results with largest score $s(u, v)$ to user $u$}
    $S \leftarrow ep$ \tcp{set of visited elements}
    $C \leftarrow ep$  \tcp{set for candidates}
    $W \leftarrow ep$  \tcp{dynamic list of results}
    \For{$t \leftarrow 1 \dots T_c$} {
    	$N \leftarrow$ union of neighbors at layer $l_c$ of all items in $C$ \;
    	$N \leftarrow N - S$ \tcp{pruning visited nodes}
    	$S \leftarrow S \cup N$ \tcp{mark as visited}
    	$W \leftarrow argTopK_{v \in W \cup N} s(u, v)$\;
    	$C \leftarrow W \cap N$ \tcp{new candidates}
    }
	\KwRet{$W$}
	\label{algorithm:dvf}
\end{algorithm}

\subsection{Search with Arbitrary Neural Metric}\label{sec:neural_metric}
\subsubsection{Motivation}
When facing arbitrarily models, triangle inequality, symmetry and convexity can no longer be exploited to validate the rationality of similarity graph search with $s_u(\ev_v)$. In practice, the reaction of $s_u(\ev_v)$ to small perturbation of $\ev_v$ is highly uncertain, e.g., $s_u(e_v)$ may fluctuate drastically when $\ev_v$ is slightly perturbed. Intuitively, this uncertainty plagues the retrieval perforamnce especially when the similarity metrics used in graph construction stage (l2 distance among $\ev_v$) and retrieval stage ($s_u(\ev_v)$) are highly heterogeneous, shown in~\Cref{tab:model_structure}. And in this work, we show that retrieval performance can be empirically augmented if we intentionally bias $s_u(\ev_v)$ to avoid uncertainty w.r.t $\ev_v$. 

Our philosophy is based upon an analogy to identifying the local optimum of a differentiable function $f:\mathbb{R}^d \rightarrow \mathbb{R}$ along with a certain direction. Suppose that the solution to $\min - s_u(e_v)$ is an arbitrary vector defined in $R^d$, gradient descent and coordinate descent are commonly used to find the local optimum. And we claim that the graph search is analogous to block coordinate descent, of which the update direction is governed by graph structure and top-k procedure instead of gradients. Hence, given the above, we can interpret the uncertainty of $s_u(\ev_v)$ w.r.t $\ev_v$ as analogous to the flat/sharpness of loss landscape in gradient-based optimization. Although disputable, it is widely thought that "flat minimal" usually generalize better compared to "sharp minimal"~\cite{yao2018hessian, li2017visualizing} because of their robustness to small perturbation of inputs. Earlier works have attempted to change the optimization algorithm to favor flat minimal and find "better" regions~\cite{hochreiter1997flat, chaudhari2019entropy, desjardins2015natural}. Inspired by these works, we leverage the adversarial training~\cite{yao2018hessian, goodfellow2014explaining, yuan2019adversarial, sinha2018gradient, shrivastava2017learning} to both mitigate the uncertainty and improve the robustness of arbitrary $s_u(\ev_v)$ w.r.t $\ev_v$.

\subsubsection{Adversarial Gradient Method} 
Generally speaking, we resort to the adversarial gradient method and introduce flatness into $s_u(\ev_{v})$ in an end-to-end learning-based method~\cite{yao2018hessian}. 

To achieve the robustness of deep neural networks by the defense against adversarial examples has been widely applied to various computer vision tasks in recent years \cite{goodfellow2014explaining, yuan2019adversarial, sinha2018gradient, shrivastava2017learning}. Adversarial examples refer to normal inputs with crafted perturbations which are usually human-imperceptible but can fool deep neural networks maliciously. The adversarial training utilized in our work is one of the most effective approaches \cite{shafahi2019adversarial, wang2021convergence} defending against adversarial examples for deep learning. More specifically, we flatten the landscape of $s_u(\ev_{v})$ w.r.t. $\ev_v$ via training on adversarially perturbed $\tilde{\ev_v}$.

In our case, our solutions to maximize $s_u(\ev_{v})$ are limited to corpus $\mathcal{V}$. Hence, we mainly focus on the landscape of $s_u(.)$ around each $\ev_v$ instead of the overall landscape. We formulate the training objective in terms of flatness as follow:

\begin{equation}\label{eq:aux_loss}
\begin{aligned}
\mathcal{L}_{AUX} &= \sum_u \sum_{v \in \mathcal{Y}_u} s_u(\ev_v)  \log \frac{s_u(\ev_v)}{s_u(\tilde{\ev_v})} \\ 
\tilde{\ev_v} &= \ev_v + \Delta
\end{aligned}
\end{equation}
where $\mathcal{Y}_u$ consists of the labels from both true distribution and noise distribution for each $u \in \mathcal{U}$ according to NCE.

As described in~\Cref{eq:aux_loss}, the flatness is translated into a trainable objective: if $s_u(\ev_v)$ has a flat landscape w.r.t $\ev_v$, it should be robust, even invariant, to small perturbations around $\ev_{v}$. Here, we use the Kullback–Leibler divergence to penalize the discrepancy between $s_u(\ev_v)$ and $s_u(\tilde{\ev_v})$ as $s_u(.)$ stands for the probability of user $u$ engages with item $v$. As mentioned in~\Cref{eq:loss_all}, the $\mathcal{L}_{AUX}$ combined with $\mathcal{L}_{NCE}$ constitute the final loss $\mathcal{L}_{all}$. The key to the auxiliary task is the direction and magnitude of perturbation $\Delta$, where the adversarial gradient method comes into play.

In detail, we generate the adversarial examples by fast gradient sign method (FGSM) \cite{goodfellow2014explaining}, which computes the perturbation as:
\begin{equation}
    \Delta = \epsilon \text{sign}(\nabla_{\ev_v}s_u(\ev_v))
\end{equation}
where the $\nabla_{\ev_v}s_u(\ev_v)$ stands for the gradient of $s_u(\ev_v)$ w.r.t. $\ev_v$ that can be easily computed by backpropagation and the max-norm of perturbation $\Delta$ is bounded by $\epsilon$.

Put simply, we achieve the search with the arbitrary measure without utilizing the convexity of $s_u(\ev_v)$. Instead, our framework is built upon the flatness of $s_u(\ev_v)$  w.r.t each $\ev_v$, which can be achieved with a simple yet effective auxiliary task.

\section{system implementation}\label{sec:sys}

\begin{figure}
    \vspace{-5pt}
    \includegraphics[width=0.5\textwidth]{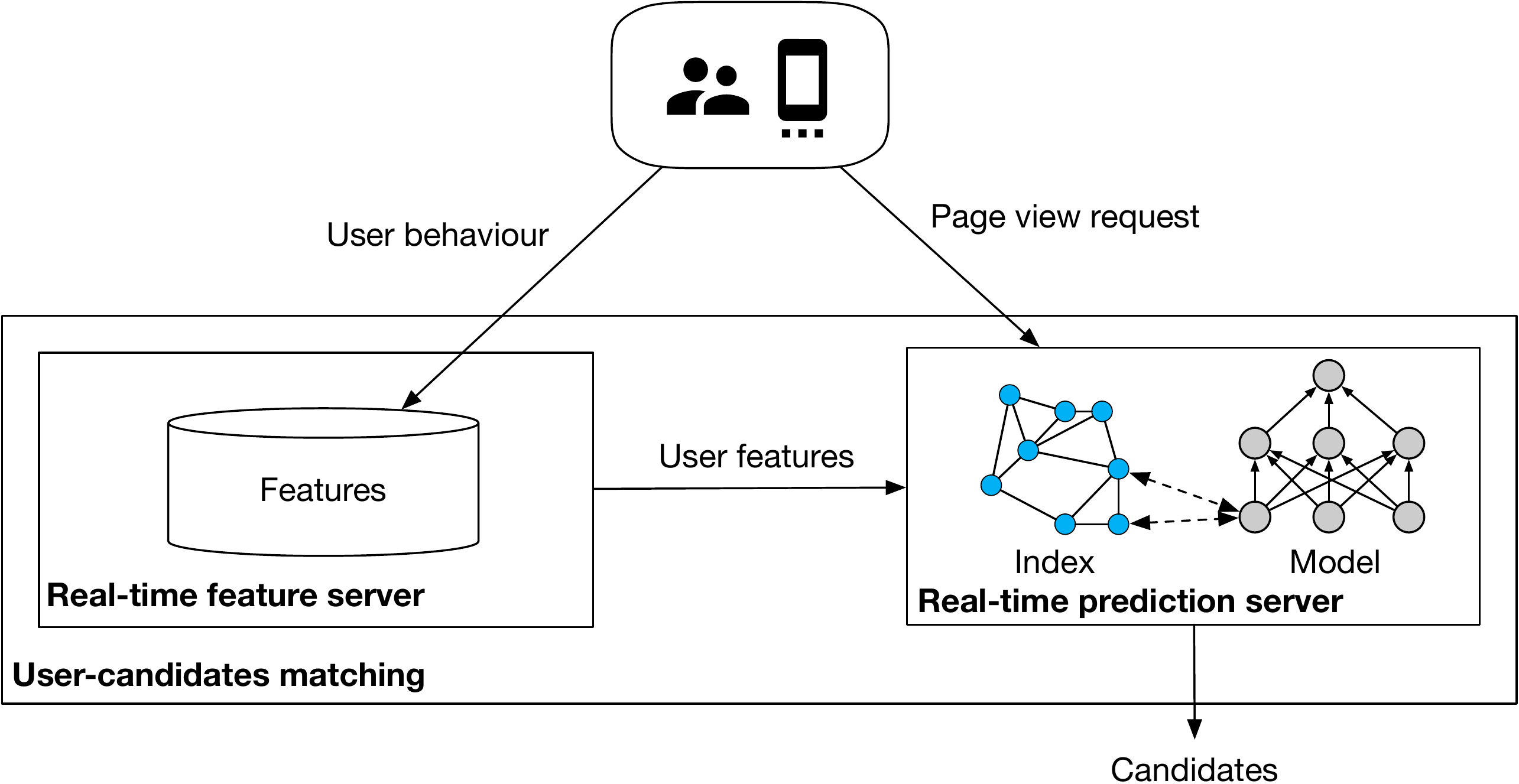}
    \vspace{-20pt}
    \caption{Online Serving System. The NANN service is provided by real-time prediction server, where graph-based index and deep neural network constitute a unified Tensorflow graph. The NANN service receives the user features from real-time feature server and output the retrieved candidate items to downstream task directly.}
    \label{fig:system}
    \vspace{-10pt}
\end{figure}

\Cref{fig:system} illustrates the online serving architecture of the proposed method. In general, almost any off-the-shelf inference system, e.g. TensorFlow Serving, for deep neural networks can provide out-of-the-box services for NANN. The framework is flexible to use and maintain since we integrate graph-based index with the deep neural network and form a unified Tensorflow graph. The neural network inference and graph-based retrieval of NANN can thus serve as a unified module. 

As described in~\Cref{algorithm:total}, the online inference is mainly composed of feed-forward of $s_u(\ev_v)$ and search on the graph, which performs alternatively. For online computation, we place the search on the graph on the Central Processing Unit (CPU) to maintain the flexibility of retrieval, while placing feed-forward of $s_u(\ev_v)$ on the Graphics Processing Unit (GPU) for efficiency. Correspondingly, both the graph-based index and the precomputed $\ev_v$ are represented as Tensorflow tensor and cached in CPU memory. The host-to-device and device-to-host communications follow the latest Peripheral Component Interconnect Express (PCIe) bus standard. This design can achieve a balance between flexibility and efficiency while just introducing slight communication overheads.

Graph representation lays the foundation for online retrieval. In our implementation, each $v \in \mathcal{V}$ is firstly serially numbered and assigned with a unique identifier. The hierarchical structure of HNSW is then represented by multiple Tensorflow RaggedTensors~\footnote{https://www.tensorflow.org/guide/ragged\_tensor}. 

Here, we mainly emphasize the online serving efficiency optimizations of our proposed method, which are based on the Tensorflow framework.

\subsection{Mark with Bitmap}
To ensure the online retrieval performance, it is of importance to increase the outreach of candidate items within limited rounds of neighborhood propagation, as shown in~\Cref{algorithm:dvf}. Hence, we need to mark the visited items and bypass them to traverse further. The idea of Bitmap comes to mind as the $v$ is serially numbered. We invent the Bitmap procedure by building C++ custom operators (Ops) within the Tensorflow framework. We summarize the performance of Bitmap Ops in terms of queries per second (QPS) and response time (RT) in milliseconds in~\Cref{tab:perf}.

\begin{table}[!htbp]
\centering
\vspace{-5pt}
\caption{Different implementations of ``Mark'' procedure}
\vspace{-5pt}
\label{tab:perf}
  \begin{threeparttable}
\begin{tabular}{c c c c c}
\toprule
$\{ ef_2, ef_1, ef_0 \}$                            & $\{ T_2, T_1, T_0 \}$                      & Ops    & QPS & RT (ms) \\ 
\hline
\multirow{2}{*}{$\{ 200, 500, 1000 \}$} & \multirow{2}{*}{$\{ 1, 1, 2 \}$} & Raw    & 185 & 23.4                \\
&                     & Bitmap & \textbf{624} & \textbf{6.3}                 \\
\hline
\end{tabular}
\begin{tablenotes}
      \scriptsize	
      \item  1 GPU (Nvidia T4); 32 CPU cores (Intel(R) Xeon(R) Platinum 8163).
      \item  Deep neural network is accelerated by half-precision and XLA.
    \end{tablenotes}
  \end{threeparttable}
\vspace{-10pt}
\end{table}

We test the performance of Bitmap Ops with the model architecture of $s_u(\ev_v)$ deployed in production, of which detailed configuration will be introduced in~\Cref{sec:exp}. We traverse a three-layer graph-based index with the $|\mathcal{V}|$ equal to 1,300,000 and tune the parameters in~\Cref{algorithm:dvf} to control the number of candidates, roughly 17,000 for the benchmark testing, to be evaluated. As demonstrated ~\Cref{tab:perf}, our custom Bitmap Ops significantly outperform the Tensorflow Raw Set Ops.

\subsection{Dynamic Shape with XLA}

XLA (Accelerated Linear Algebra) is a domain-specific compiler for linear algebra that can accelerate the TensorFlow model~\footnote{https://www.tensorflow.org/xla}. XLA can automatically optimize the model execution in terms of speed and memory usage by fusing the individual Tensorflow Ops into coarsen-grained clusters. Our model has achieved a \char`\~ 3x performance improvement with the help of XLA. However, it requires all tensors of the computation graph to have fixed shapes and compiled codes are specialized to concrete shapes. In our scenario, the number of unvisited items $|C|$ to be evaluated by $s_u(\ev_v)$ is dynamic for each neighborhood propagation in~\Cref{algorithm:dvf}. Therefore, we present an "auto-padding" strategy to transform the dynamic shapes, e.g., $|C|$ in~\Cref{algorithm:dvf}, to certain predefined and fixed shapes. In detail, we set in advance a grid of potential shapes of $|C|$ and generate compiled codes for these predefined shapes with  XLA's Just-in-Time ( JIT ) compilation, which is triggered by replaying the logs from the production environment. For online inference, the "auto-padding" strategy automatically pad the tensor with size $|C|$ to its nearest greater point on the grid and execute efficiently with its corresponding compiled code by XLA, and slice the tensor to its original shape afterward. In short, we extend the capacity of XLA to dynamic shapes with an automatic "padding-slicing" strategy. 

\section{experiments}\label{sec:exp}
We study the performance of the proposed method as well as present the corresponding analysis in this section. Besides comparison to baseline, we put more emphasis on the retrieval performance of NANN and the corresponding ablation study due to the inadequacies of directly related works. Experiments on both an open-source benchmark dataset and an industry dataset from Taobao are conducted to demonstrate the effectiveness of the proposed method. We observe that our proposed method can significantly outperform the baseline and achieve almost the same retrieval performance as its brute-force counterpart with much fewer computations.

\subsection{Setup}

\subsubsection{Datasets} 
We do experiments with two large-scale datasets: 1) a publicly accessible user-item behavior dataset from Taobao called UserBehavior~\footnote{https://tianchi.aliyun.com/dataset/dataDetail?dataId=649}; 2) a real industry dataset of Taobao collected from traffic logs.~\Cref{tab:data} summarizes the main statistics for these two datasets.

\begin{table}[!htbp]
\centering
\vspace{-5pt}
\caption{Statistics of evaluation datasets}
\vspace{-5pt}
\label{tab:data}
\begin{tabular}{lcc}
\toprule
 & \textbf{UserBehavior} & \textbf{Industrial Data of Taobao} \\
\midrule 
\# of users & 976,779 & 100 million \\
\# of items & 4,163,442 & 1.3 million \\
\# of records & 85,384,110 & 375 million \\
\bottomrule
\end{tabular}
\vspace{-5pt}
\end{table}

\textbf{UserBehavior}. UserBehavior is a subset of Taobao user behaviors for recommendation problems with implicit feedback. Each record includes user ID, item ID, item category ID, behavior type, and timestamp. The behavior type indicates how the user interacts with the item, including click, purchase, adding items to the shopping cart, and adding items to favorites. We filter some of the users with high sparsity and keep the users with at least 10 behaviors. Suppose that the behaviors of user $u$ be $(b_{u_1},\dots,b_{u_k},\dots,b_{u_n})$, the task is to predict $b_{u_{k+1}}$ based on the preceding behaviors. The validation and test sets are constituted by the samples from randomly selected 10,000 users respectively. We take the $\ceil*{\nicefrac{l_u}{2}}$-th ($l_u$ denotes the length of behavior sequence for user $u$) behavior of each $u$ as ground truth and predict it based on all behaviors before.

\textbf{Industrial Data of Taobao}. The industry dataset is collected from the traffic logs in the Taobao platform, which is organized similarly to UserBehavior but with more features and records. The features of the industry dataset are mainly constituted by user profile, user behavior sequence, and item attributes. 

\subsubsection{Metrics}

We use $\text{recall-all}@M$,  $\text{recall-retrieval}@M$,  $\text{recall-}\Delta@M$, $\text{coverage}@M$ to evaluate the effectiveness of our proposed method. In general, for a user $u$, the recall can be defined as
$$
\text{recall}(\mathcal{P}_u, \mathcal{G}_u)@M(u) = \frac{|\mathcal{P}_u \cap \mathcal{G}_u|}{|\mathcal{G}_u|}.
$$
where $\mathcal{P}_u (|\mathcal{P}_u| = M)$ denotes the set of retrieved items and $\mathcal{G}_u$ denotes the set of ground truths.

The capacity of a trained scoring model $s(v, u)$ is assessed by exhaustively evaluating the corpus $\mathcal{V}$ for  $u \in \mathcal{U}$, namely, 
$$
\text{recall-all}@M(u) = \text{recall}(\mathcal{B}_u, \mathcal{G}_u)@M(u)
$$
 where $\mathcal{B}_u = \argTopk_{v \in \mathcal{V}} s_u(\ev_{v})$ ($|\mathcal{B}_u| = M$) is the set of precisely top-$k$ scored items that can be produced by brute-force scanning.

Suppose that we traverse the graph-based index by $s(v, u)$ and retrieve relevant items $\mathcal{R}_u$ ($|\mathcal{R}_u|=M$|) for each user $u$, the retrieval recall then can be evaluated by,
$$
\text{recall-retrieval}@M(u)= \text{recall}(\mathcal{R}_u, \mathcal{G}_u)@M(u)
$$

Correspondingly, the retrieval loss in terms of recall introduced by graph-based index can be defined as,
$$
\text{recall-}\Delta@M(u) = \frac{\text{recall-all}@M - \text{recall-retrieval}@M}{\text{recall-all}@M}.
$$

Furthermore, we make use of $\text{coverage}@M(u)$ to describe the discrepancy between the brute-force scanning and the retrieval. Formally, 
$$
 \text{coverage}@M(u) = \frac{|\mathcal{R}_u \cap \mathcal{B}_u|}{|\mathcal{B}_u|}
$$

From now on, we refer to the retrieval quality as the consistency between the items from retrieval and those from the brute-force, measured by $\text{recall-}\Delta@M$ and $\text{coverage}@M$.

Finally, we take the average over each $u$ to obtain the final metrics, where $u$ is from the testing set.

\subsubsection{Model architecture}
The model architecture (denoted as DNN w/ attention) is illustrated in~\Cref{fig:network}, which contains user network, target attention network, item network, and score network. More details are in Appendix.
To measure the model capacity and retrieval performance of different model structures, we also conduct experiments on the following model structures:
1) DNN w/o attention, which replaces the target attention network with a simple sum-pooling over the embeddings of user behavior sequence; 2) two-sided, which only consists of user embedding (the concatenation of the output of user network and the sum-pooling over the embeddings of user behavior sequence) and item embedding, and calculate the user-item preference score by inner product.

\subsubsection{Implementation details}

\begin{figure*}[!htb]
    \centering
    \vspace{-5pt}
    \subfigure[Recall-Industry]{
        \includegraphics[width=0.23\textwidth]{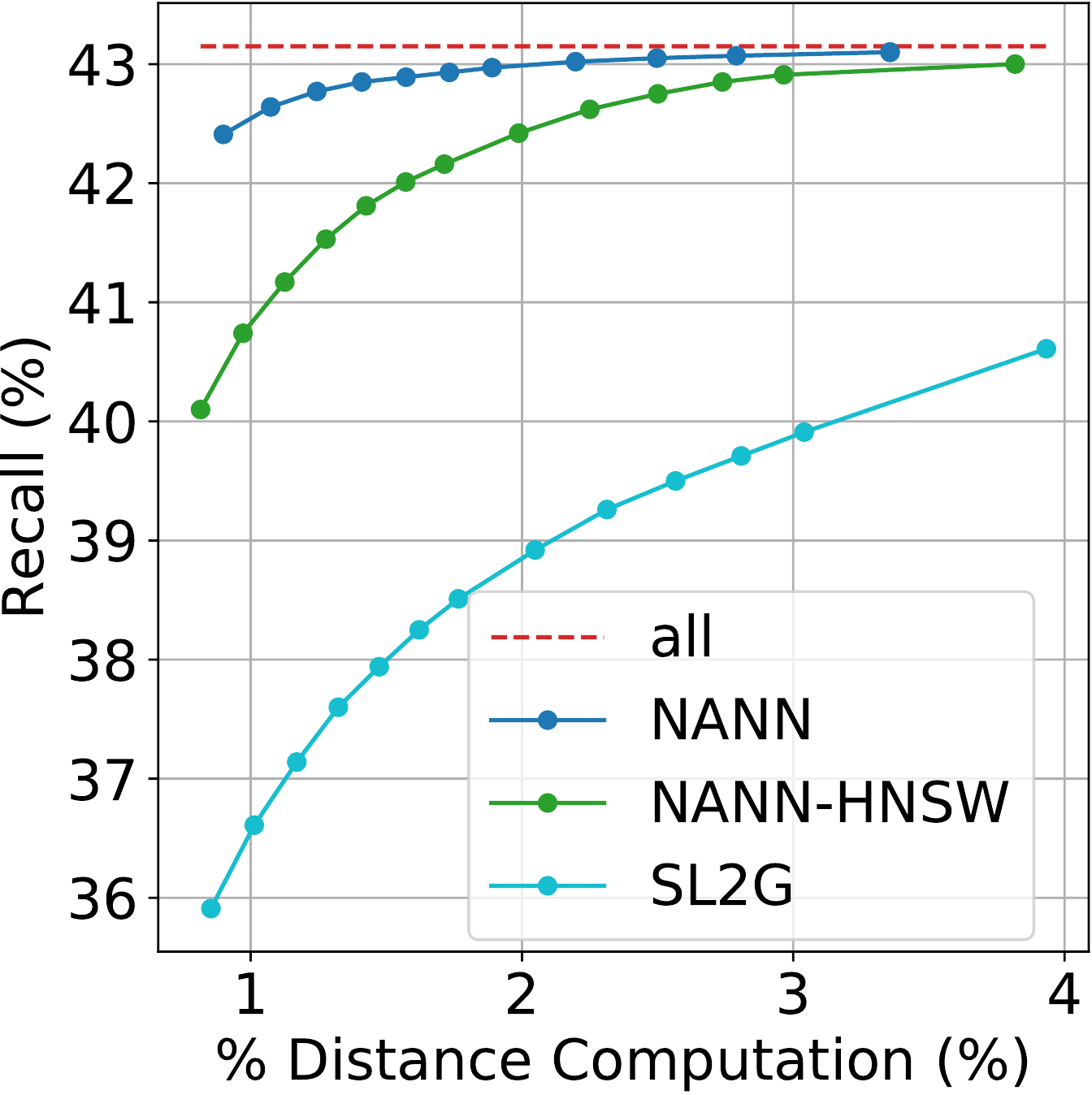}
        \label{fig:main_recall_industry}
    }
	\hfil
    \subfigure[Coverage-Industry]{
        \includegraphics[width=0.23\textwidth]{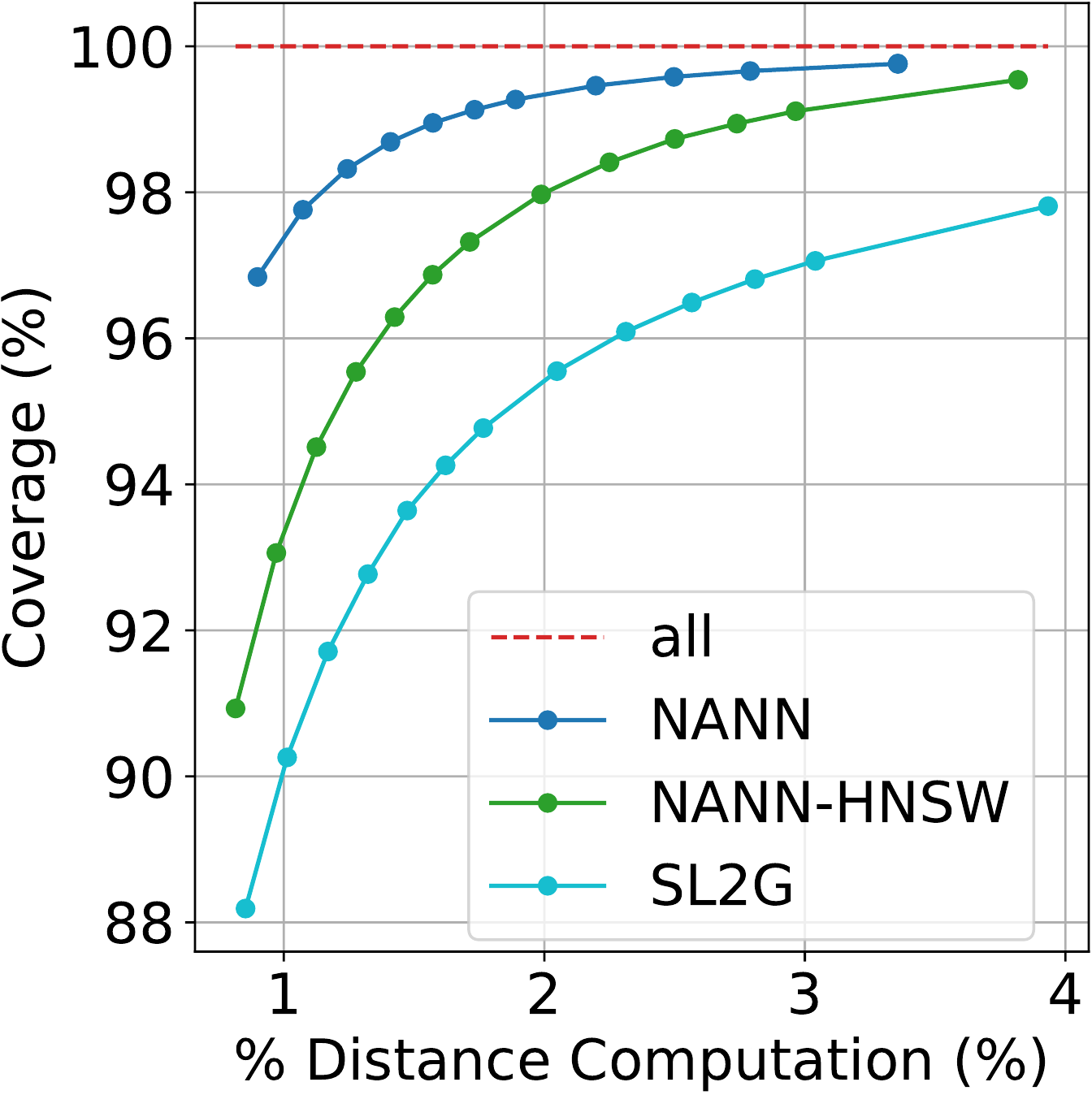}
        \label{fig:main_coverage_industry}
    }
	\hfil
    \subfigure[Recall-UserBehavior]{
        \includegraphics[width=0.23\textwidth]{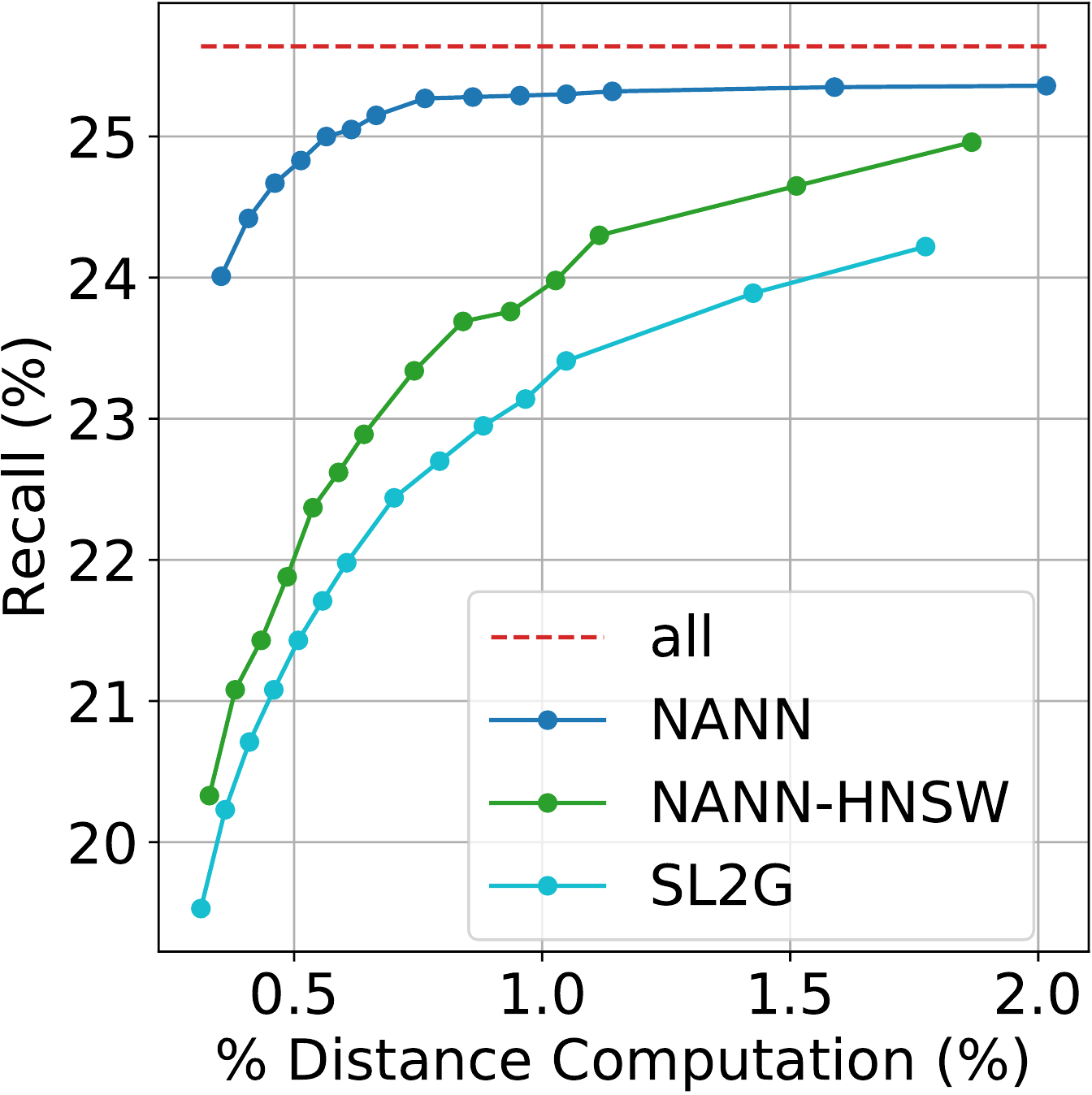}
        \label{fig:main_recall_ub}
    }
	\hfil
    \subfigure[Coverage-UserBehavior]{
        \includegraphics[width=0.23\textwidth]{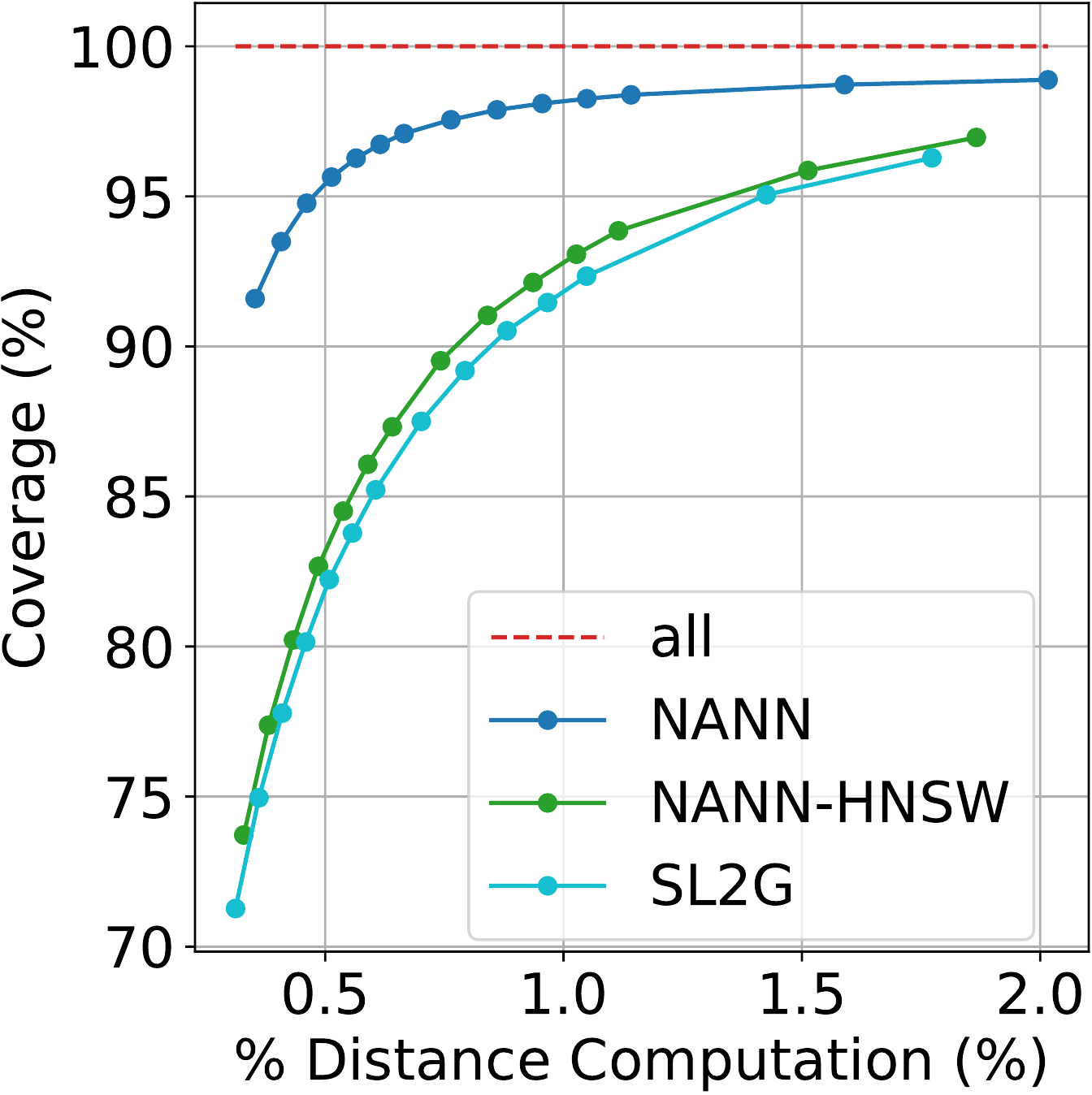}
        \label{fig:main_coverage_ub}
    }
    \vspace{-10pt}
    \caption{Results of our proposed NANN and SL2G on Industry and UserBehavior dataset.}
    \vspace{-5pt}
    \label{fig:main_results}
\end{figure*}

Given the dataset and model structure, we train the model with the loss function defined in~\Cref{eq:loss_all}. Adam optimizer with learning rate 3e-3 is adopted to minimize the loss. The $\epsilon$ of FGSM is set to 1e-2 for the industry dataset and 3e-4 for the UserBehavior dataset. We optimize the models by NCE and assign each label from the true distribution with 19 and 199 labels from noise distribution for the industry dataset and UserBehavior dataset respectively.

After training, we extract the item feature after the item network for all valid items to build the HNSW graph. The standard index build algorithm~\cite{malkov2018efficient} is used, the number of established connections is set to 32, size of the dynamic candidate list in the graph construction stage is set to 40.

In the retrieval stage, we exhaustively calculate the scores of items in layer 2, which consists of millesimal items of entire vocabulary and can be scored in one batch efficiently. Then top-k relevant items, with k=$ef_2$, are retrieved as enter points for the following retrieval. The default retrieval parameter is set as $\{ ef_2, ef_1, ef_0 \} = \{100, 200, 400\}, \{ T_2, T_1, T_0 \} = \{1, 1, 3\}$, described in~\Cref{algorithm:total,algorithm:dvf}. Without further claim, we report top-200 ($M=200$) metrics for final retrieved items.

All the hyper-parameter are determined by cross-validation.

\begin{figure*}[!thbp]
    \centering
    \vspace{-5pt}
    \subfigure[Recall-Industry]{
        \includegraphics[width=0.23\textwidth]{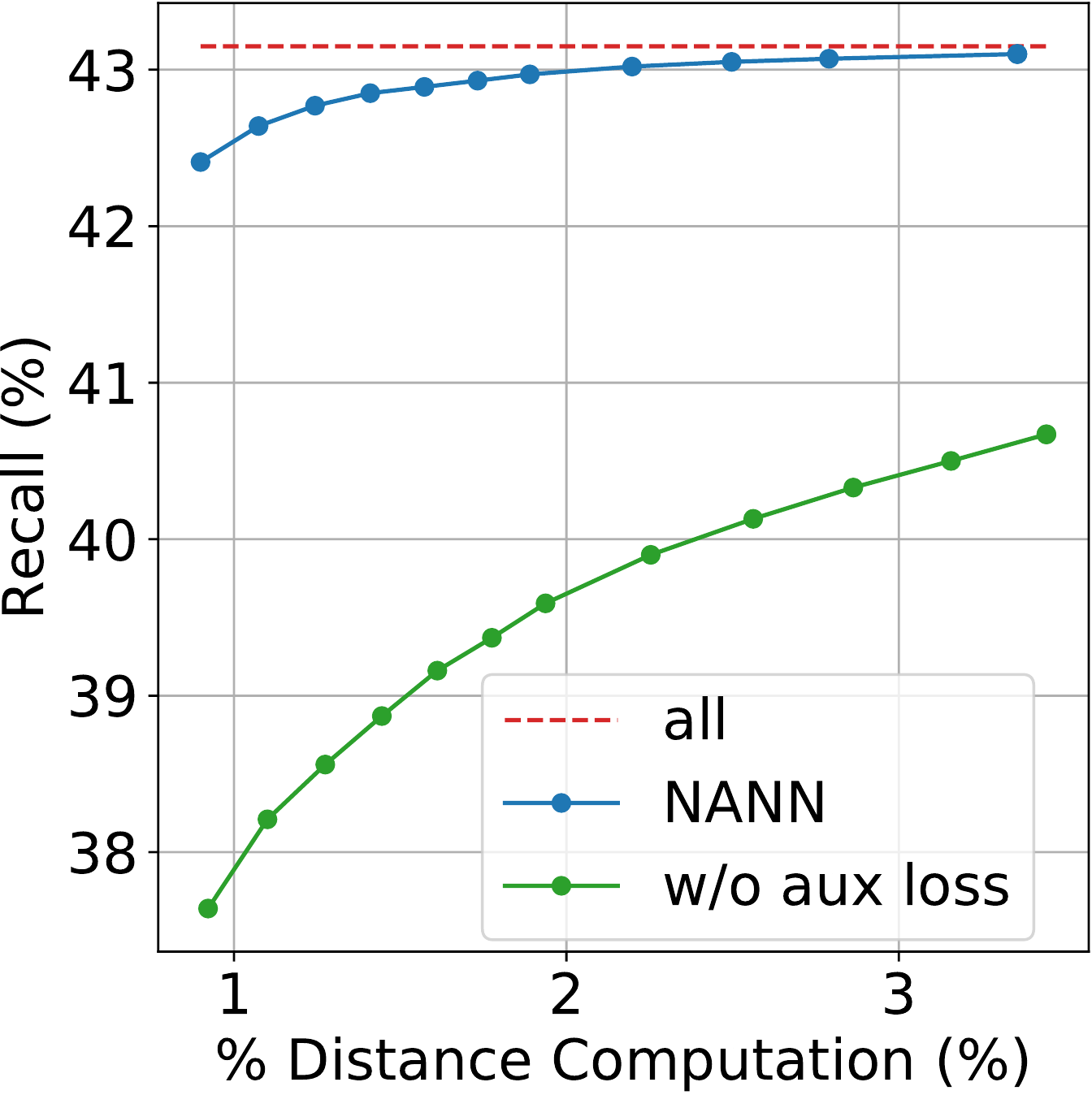}
        \label{fig:computation_recall_industry}
    }
	\hfil
    \subfigure[Coverage-Industry]{
        \includegraphics[width=0.23\textwidth]{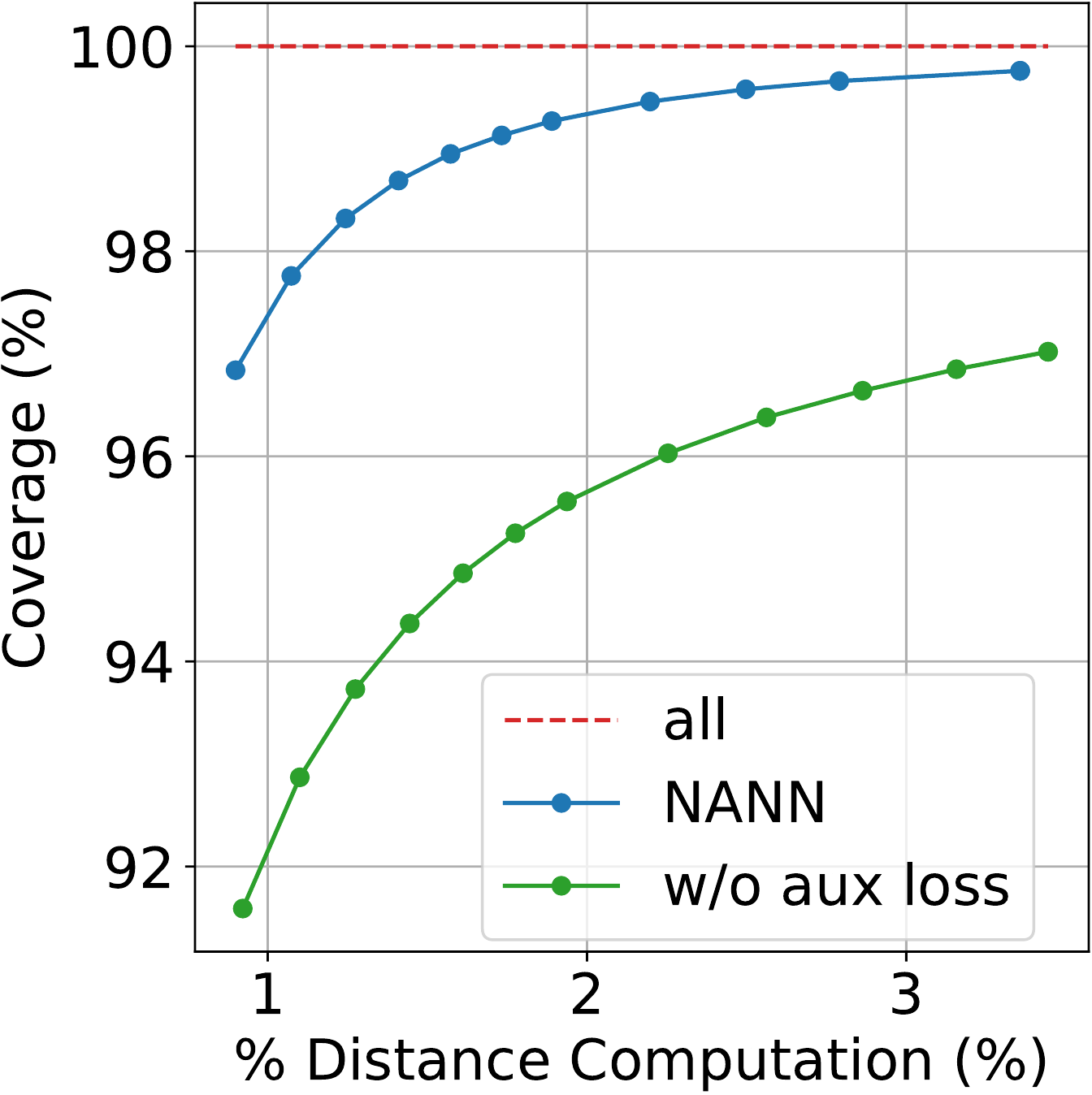}
        \label{fig:computation_coverage_industry}
    }
	\hfil
    \subfigure[Recall-UserBehavior]{
        \includegraphics[width=0.23\textwidth]{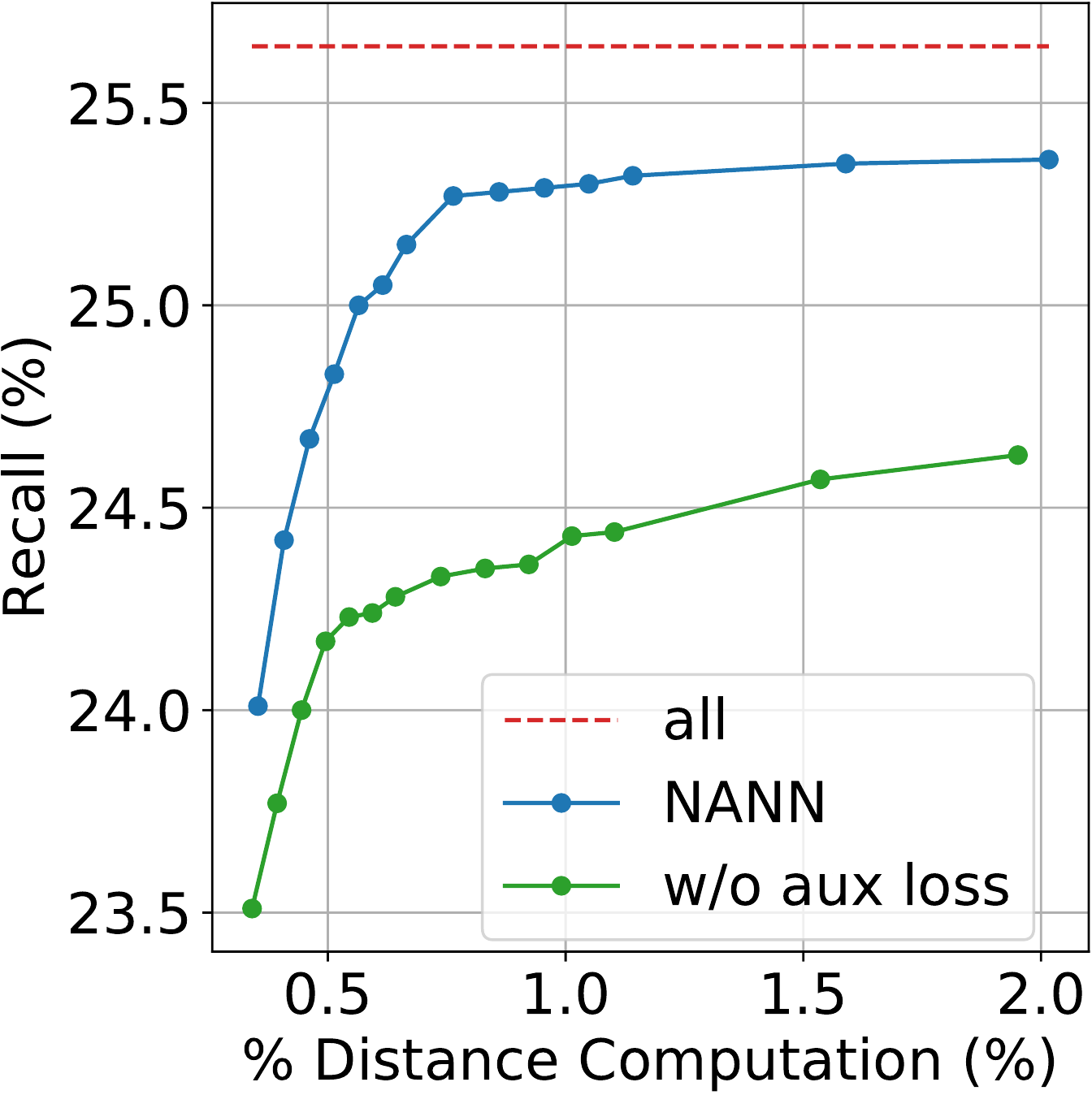}
        \label{fig:computation_recall_ub}
    }
	\hfil
    \subfigure[Coverage-UserBehavior]{
        \includegraphics[width=0.23\textwidth]{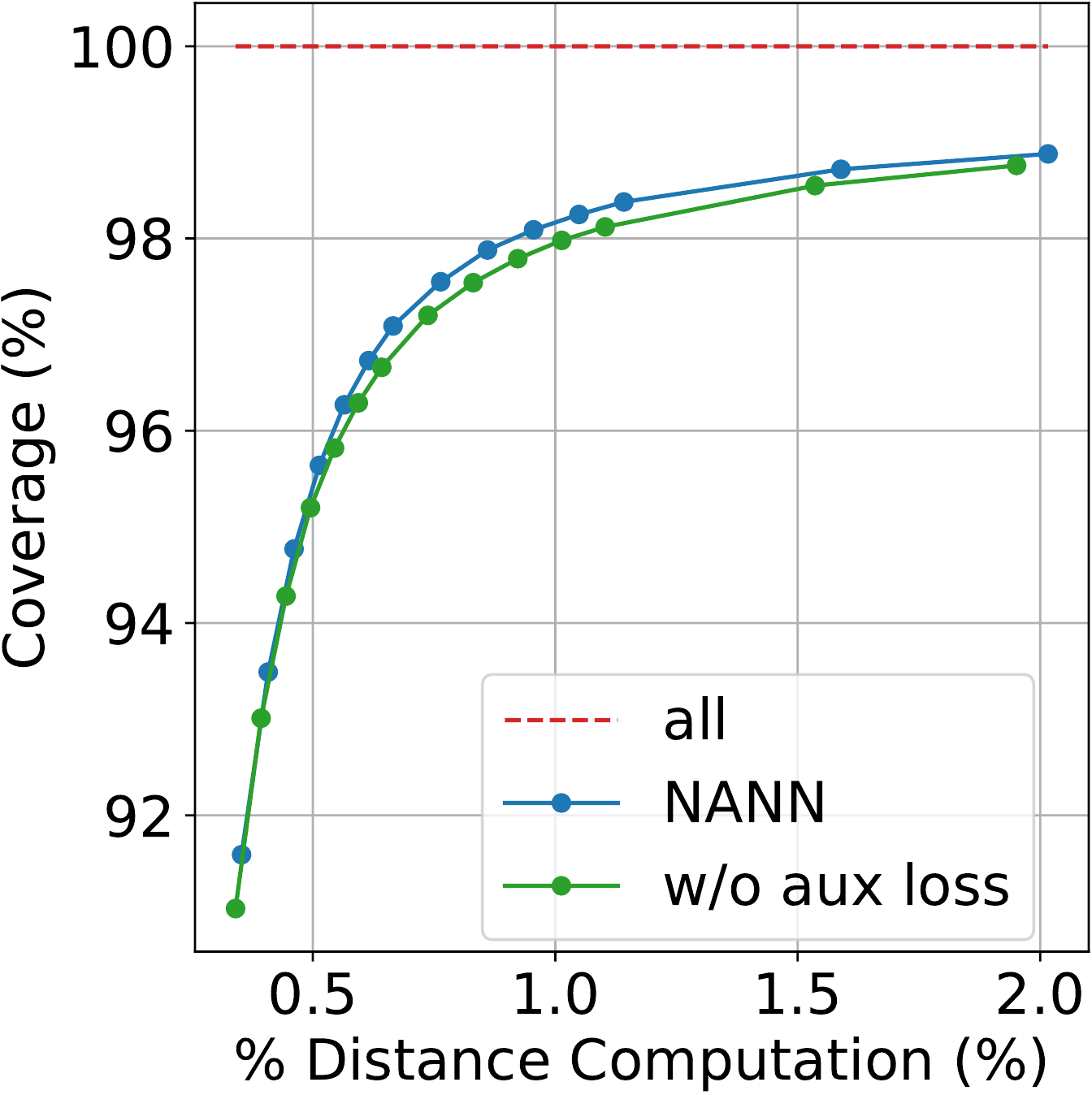}
        \label{fig:computation_coverage_ub}
    }
    \vspace{-10pt}
    \caption{The effect of adversarial gradient training on Industry and UserBehavior dataset.}
    \label{fig:adv_results}
    \vspace{-5pt}
\end{figure*}

\subsection{Results}

\subsubsection{Comparison to Baselines}

We compare with the baseline method SL2G, which directly leverages HNSW-retrieval with the deep model in the HNSW graph constructed by l2 distance among $\ev_v$. The comparison results of different methods are shown in~\Cref{fig:main_results}. Each x-axis stands for the ratio of the number of traversed items to $|\mathcal{V}|$ for reaching the final candidates. 

First of all, NANN achieves great improvements on recall and coverage in comparison with SL2G across different numbers of traversed items for the two datasets. Especially, NANN outperforms SL2G by a larger margin when we evaluate a smaller portion of items to reach the final items.

Second, NANN performs on par with its brute-force counterpart by much fewer computations. Especially, NANN hardly plagues retrieval quality and achieves 0.60\% recall-$\Delta$ and 99.0\% coverage with default retrieval parameter when applied to industrial data of Taobao. Moreover, the model capacity and robustness indicated by $\text{recall-all}$ can also benefit from the defense against moderate adversarial attacks. 

Finally, NANN can rapidly converge, in terms of traversed items, to a promising retrieval performance. As described by the curvatures of~\Cref{fig:main_results}, only 1\% \char`\~  2\% of $\mathcal{V}$ need to be evaluated to reach a satisfying retrieval quality for the two datasets.  

\subsubsection{Beam-retrieval vs HNSW-retrieval}  \Cref{fig:main_results} demonstrate the recall and coverage for the Beam-retrieval (the ``NANN'' curve) and the original HNSW-retrieval (the ``NANN-HNSW'' curve) respectively. As shown in these figures,~\Cref{algorithm:dvf} outperforms the HNSW-retrieval version in two ways: 1) it performs consistently better across different numbers of traversed items; 2) it converges to the promising retrieval quality more rapidly. Moreover, as shown in~\Cref{fig:rounds_results}, the while-loop of HNSW-retrieval results in redundant rounds of neighborhood propagation in the ground layer which is unnecessary for recall and coverage.

\begin{figure*}[!thbp]
    \centering
    \vspace{-5pt}
    \subfigure[Recall-Industry]{
        \includegraphics[width=0.23\textwidth]{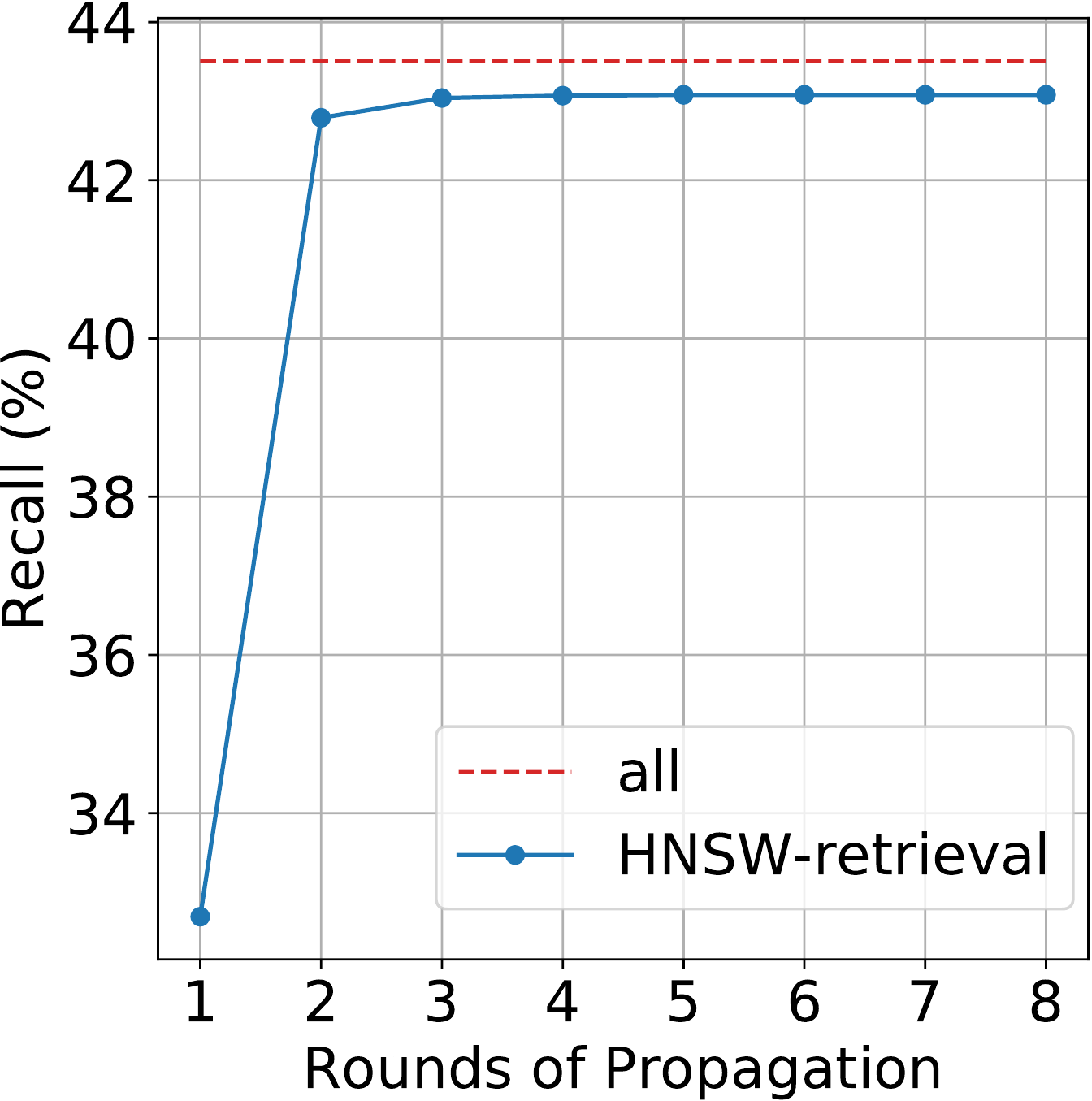}
        \label{fig:rounds_recall_industry}
    }
	\hfil
    \subfigure[Coverage-Industry]{
        \includegraphics[width=0.23\textwidth]{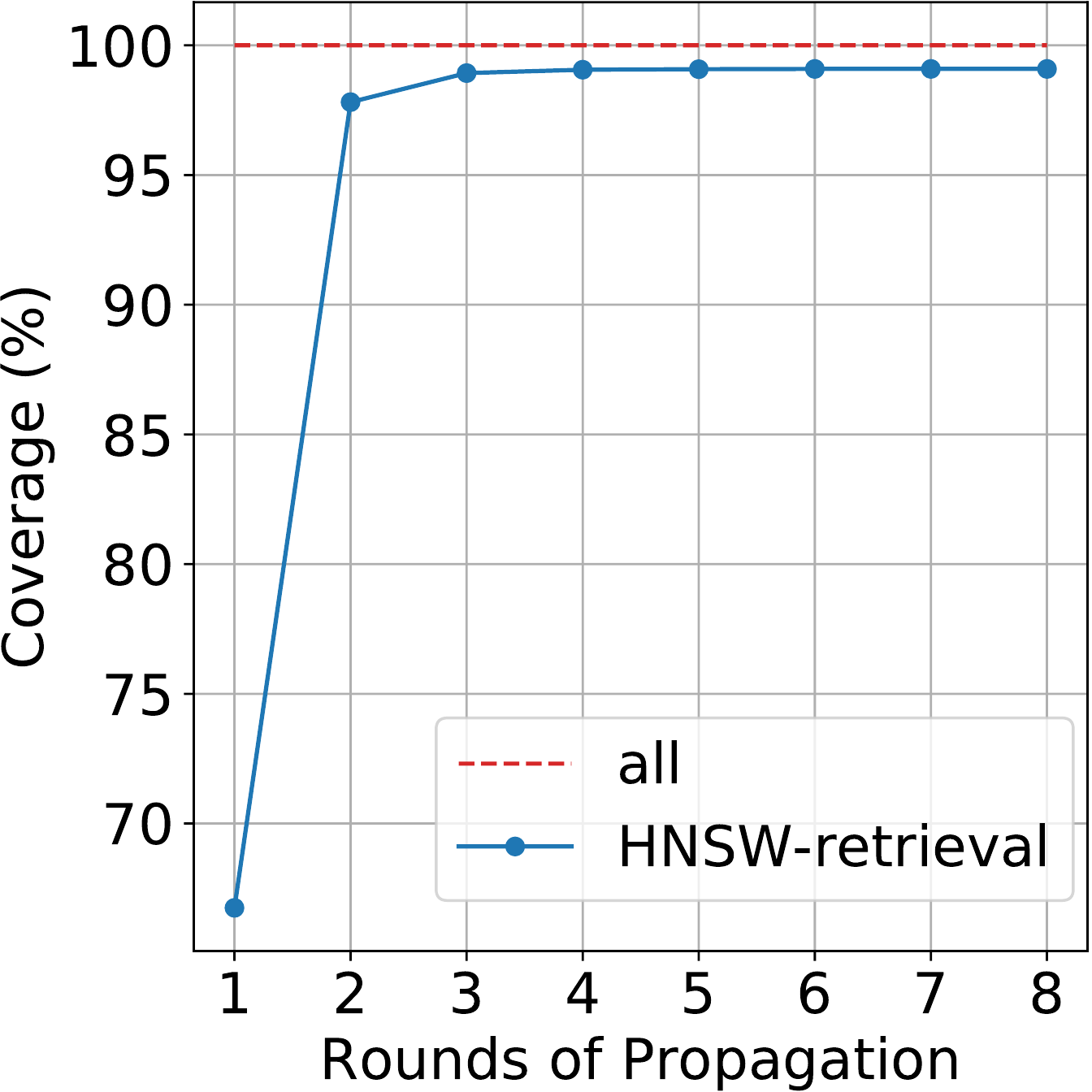}
        \label{fig:rounds_coverage_industry}
    }
	\hfil
    \subfigure[Recall-UserBehavior]{
        \includegraphics[width=0.23\textwidth]{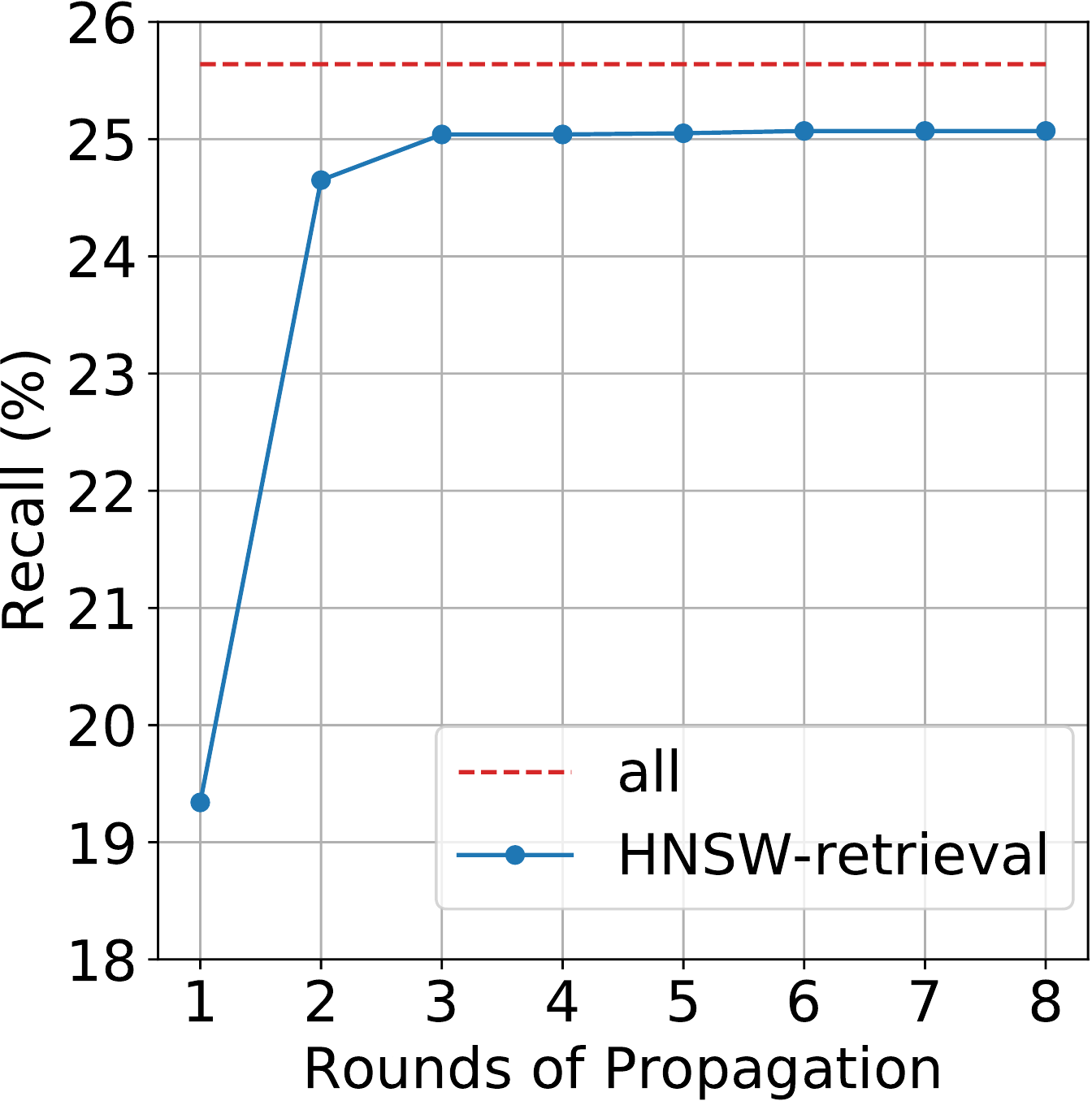}
        \label{fig:rounds_recall_ub}
    }
	\hfil
    \subfigure[Coverage-UserBehavior]{
        \includegraphics[width=0.23\textwidth]{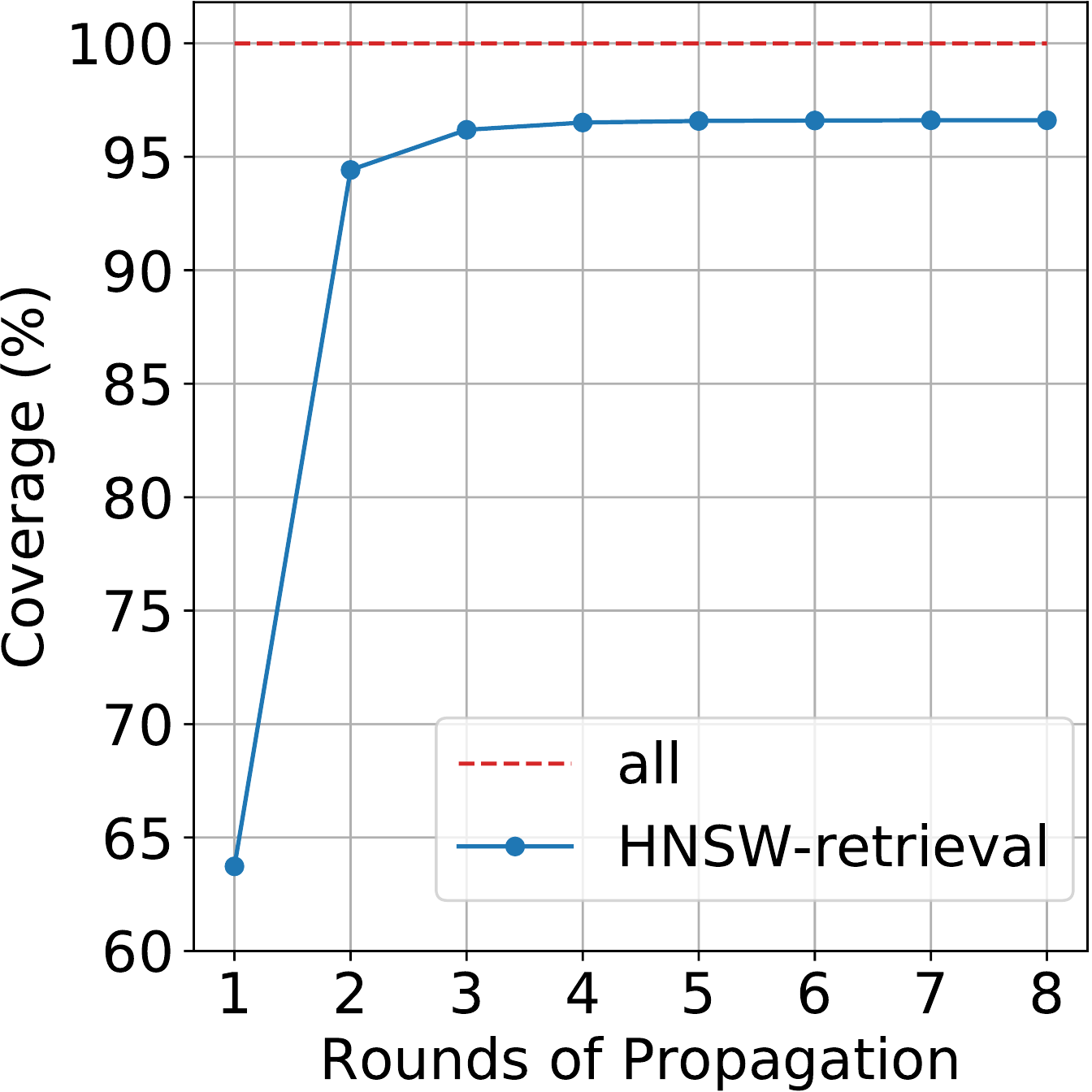}
        \label{fig:rounds_coverage_ub}
    }
    \vspace{-10pt}
    \caption{Neighborhood propagation in ground layer}
    \label{fig:rounds_results}
    \vspace{-5pt}
\end{figure*}

\subsubsection{Effectiveness of adversarial gradient training}
\begin{table}[!htbp]
\vspace{-5pt}
\setlength{\tabcolsep}{4pt}
\centering
\caption{Results of different model architectures on industry and UserBehavior dataset.}
\vspace{-5pt}
\label{tab:model_structure}
\begin{tabular}{c | c r c c c}
\toprule
\multirow{2}{*}{Dataset} & Aux & \multirow{2}{*}{model}  &  recall- & recall- & recall- \\ 
 &  loss &   &  retrieval & all & $\Delta$ \\ 
\hline
\multirow{6}{*}{Industry}
& \multirow{3}{*}{w/o} & two-sided              & 28.8\% & 28.9\% & 0.35\% \\
&                           & DNN w/o attention & 33.9\% & 34.1\% & 0.59\% \\
&                           & DNN w/ attention  & 39.2\% & 42.8\% & 8.55\% \\ 
\cline{2-6}
& \multirow{3}{*}{w/} & two-sided                   & 30.1\% & 30.2\% & \bf{0.33\%} \\
&                               & DNN w/o attention & 34.1\% & 34.2\% & \bf{0.29\%} \\
&                               & DNN w/ attention  & \bf{42.9\%} & \bf{43.2\%} & \bf{0.60\%} \\ 
\hline
& \multirow{3}{*}{w/o} & two-sided                   & 11.3\% & 11.4 \% & 0.74\% \\
&                           & DNN w/o attention.     & 12.8\% & 13.1\% & 2.30\% \\
User &                           & DNN w/ attention  & 23.9\% & 24.7\% & 3.48\% \\ 
\cline{2-6}
Behavior & \multirow{3}{*}{w/} & two-sided          & 12.4\% & 12.5\% & \bf{0.40\%} \\
&                               & DNN w/o attention & 13.1\% & 13.3\% & 1.20\% \\
&                               & DNN w/attention   & \bf{24.9\%} & \bf{25.6\%} & 3.00\% \\ 
\bottomrule
\end{tabular}
\vspace{-10pt}
\end{table}

In \Cref{fig:adv_results}, we traverse the similarity graph with Beam-retrieval and demonstrate the effectiveness of the defense against adversarial attacks. We observe that NANN is constantly superior to the model without adversarial training across the different degrees of traversal.  

We also investigate the effects of FGSM on different model architectures. As indicated by $\text{recall-all}$ and recall-$\Delta$ in~\Cref{tab:model_structure}, we empirically show that more complex models usually generalize better and achieve higher performances but may deteriorate the retrieval quality. Based on this observation, we claim that the growing discrepancy between $\text{recall-all}$ and $\text{recall-retrieval}$ may stem from the higher heterogeneity between similarity measures, and thus exploit adversarial training to mitigate the discrepancy. The default retrieval parameter is used for the comparison. As shown in \Cref{tab:model_structure}, the performance of all model architectures ranging from simple to complex can benefit from the adversarial training; FGSM can greatly improve the retrieval quality, especially for more complex models.

\subsubsection{Analysis for adversarial gradient training}
\begin{figure}
	\vspace{-5pt}
    \centering
    \includegraphics[width=0.9\linewidth]{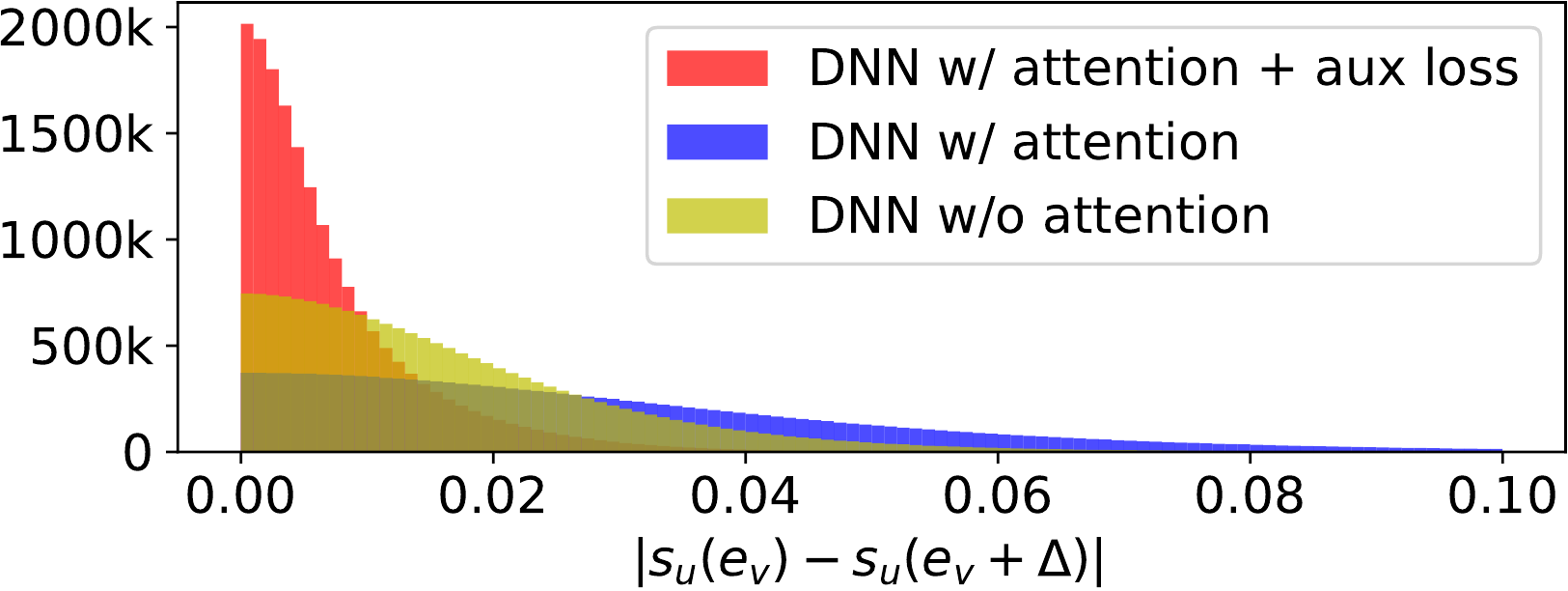}
    \vspace{-5pt}
    \caption{Reaction to small perturbations on the Industry dataset.}
    \label{fig:perturbation}
    \vspace{-10pt}
\end{figure}

\Cref{fig:perturbation} shows the reaction of model to adversarial attack after model training. We define the $\Delta$ of the adversarial attack as $\epsilon \cdot \text{rand}(-1, 1)$ akin to FGSM and compare the robustness of different models by visualizing $|s_u(\ev_{v})-s_u(\ev_{v}+\Delta)|$.~\Cref{fig:perturbation} is the histogram of $|s_u(\ev_{v})-s_u(\ev_{v}+\Delta)|$ where $v \in \argTopk_{v \in \mathcal{V}} s_u(\ev_{v})$. As demonstrated in~\Cref{fig:perturbation}, the retrieval quality empirically correlates to the robustness of model when faced with adversarial attack: 1) the greater right-skewed distribution of $|s_u(\ev_{v})-s_u(\ev_{v}+\Delta)|$ for model without attention demonstrates its superior robustness to model with attention, which is consistent with their recall-$\Delta$ in~\Cref{tab:model_structure}; 2) the retrieval quality of model with attention can be significantly improved by FGSM, and meanwhile its distribution of $|s_u(\ev_{v})-s_u(\ev_{v}+\Delta)|$ become more skewed to the right with adversarial training.

\subsubsection{Sensitivity analysis}

\begin{figure}[!thbp]
    \centering
    \vspace{-5pt}
    \subfigure[$\epsilon$]{
        \includegraphics[width=0.465\linewidth]{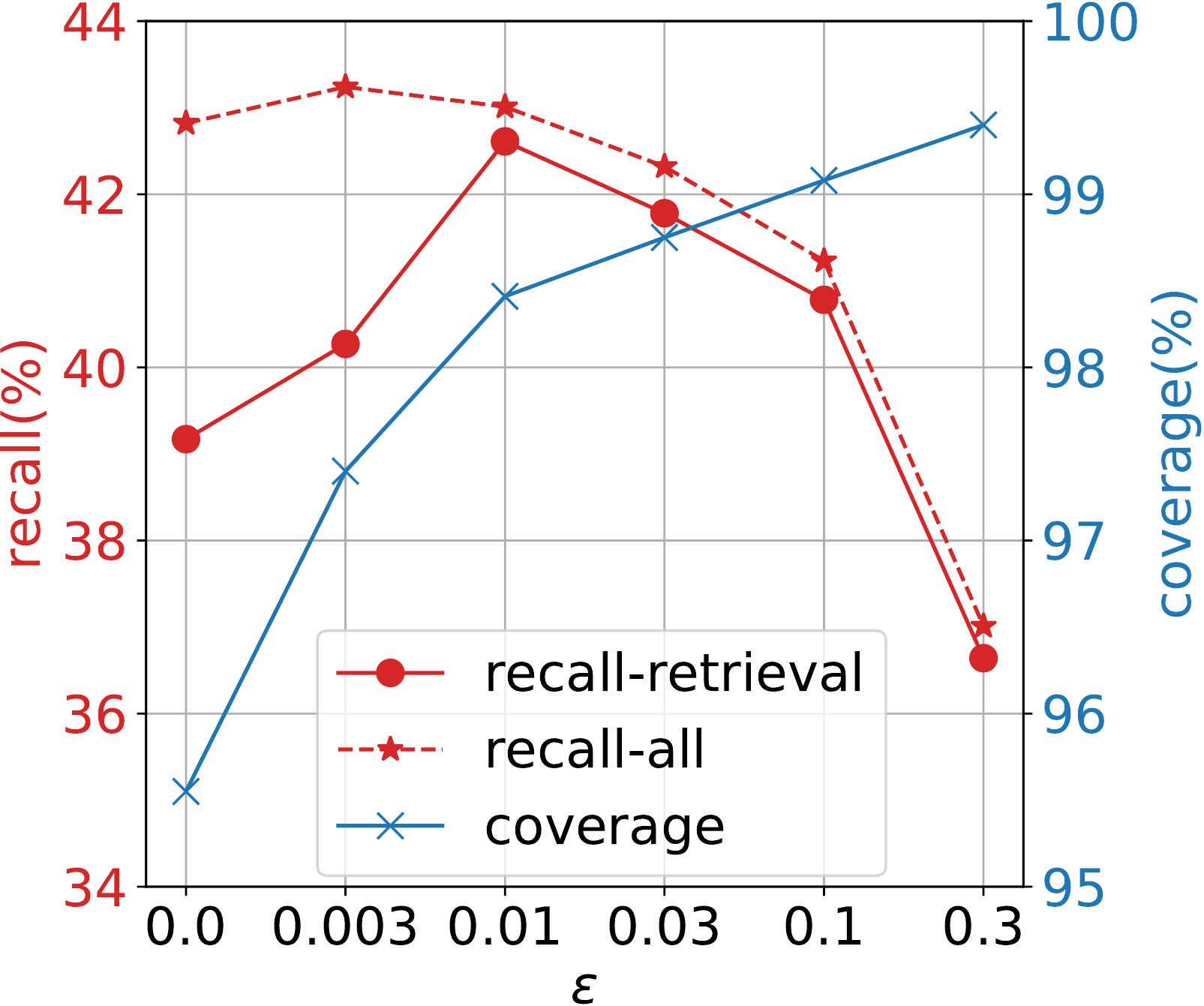}
        \label{fig:epsilon}
    }
    \subfigure[top-k]{
        \includegraphics[width=0.465\linewidth]{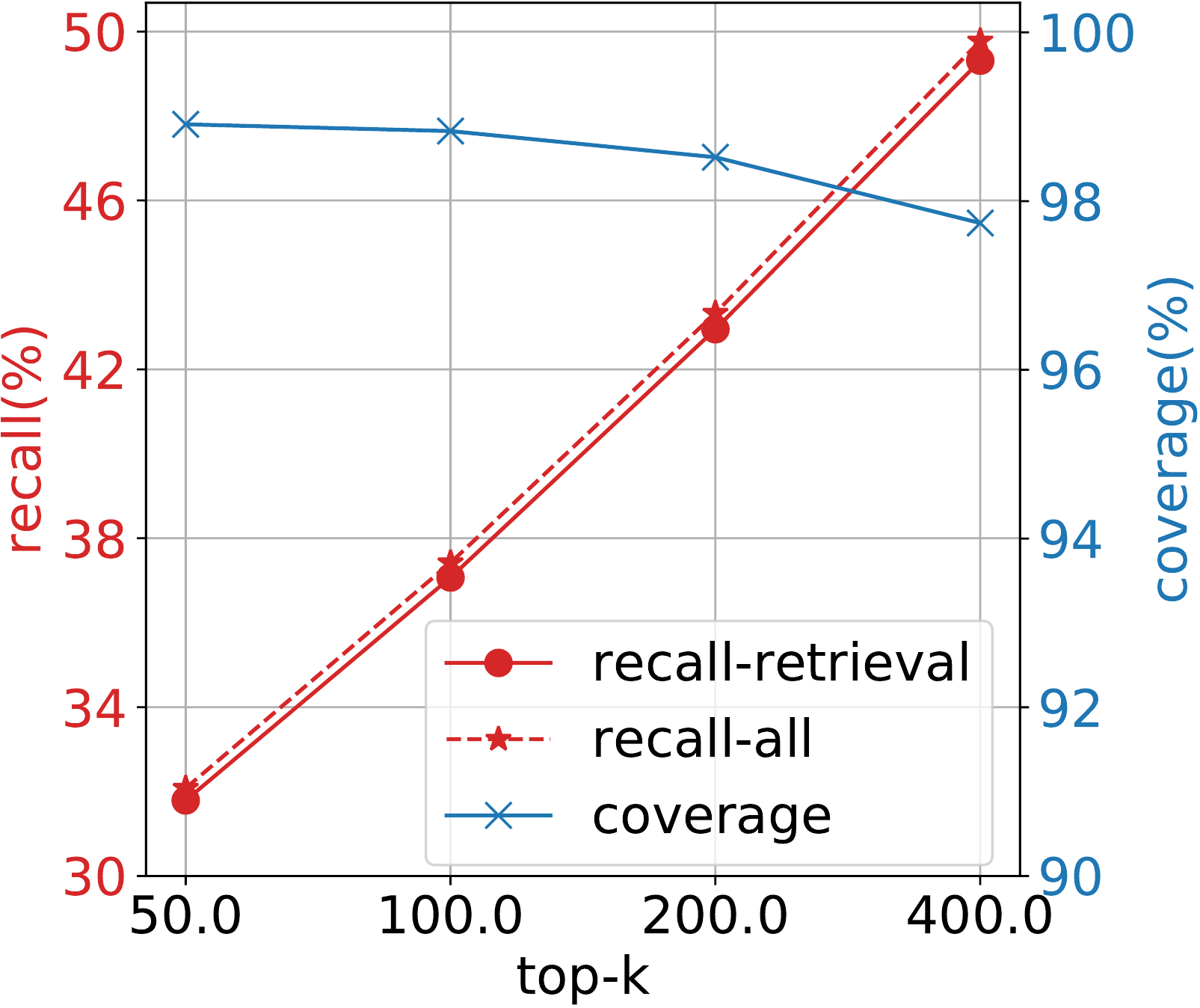}
        \label{fig:topk}
    }
    \vspace{-10pt}
    \caption{Sensitivity Analysis for $\epsilon$ in FGSM and top-k of  ground layer in Beam-retrieval on the Industry dataset.}
    \label{fig:rounds_results}
    \vspace{-10pt}
\end{figure}

\paragraph{Magnitude of $\epsilon$} \Cref{fig:epsilon} shows the correlation between $\epsilon$ and retrieval quality measured by coverage. In general, the retrieval quality is positively correlated with the magnitude of $\epsilon$. Besides, adversarial attacks can be beneficial to the overall performance measured by recall-${\text{all}}$ with a mild magnitude of $\epsilon$, but harmful when $\epsilon$ gets excessively large. Hence, the magnitude of $\epsilon$ plays an important role in the balance between retrieval quality and overall performance.

\paragraph{Different top-k}~\Cref{fig:topk} shows the effects of our proposed method on different $k$ for the final top-k retrieved items in \Cref{algorithm:total}. NANN performs consistently well across different $k$. The retrieval quality can be still guaranteed despite retrieving with larger $k$. Therefore, our method is insensitive to $k$ in general.

\subsection{Online Results}
Our proposed method is evaluated with real traffic in the Taobao display advertising platform. The online A/B experiments are conducted on main commercial pages within Taobao App, such as the "Guess What You Like" page, and last more than one month. The online baseline is the latest TDM method with Bayes optimality under beam search~\cite{zhu2018learning, zhu2019joint, zhuo2020learning}. For a fair comparison, we only substitute TDM, one of the channels in the candidate generation stage, with NANN and maintain other factors like the number of candidate items that delivered to the ranking stage unchanged. Two common metrics for online advertising are adopted to measure online performance: Click-through Rate (CTR) and Revenue per Mille (RPM).
$$
\text{CTR} = \frac{\text{\# of clicks}}{\text{\# of impressions}}, \ \text{RPM} = \frac{\text{Ad revenue}}{\text{\# of impressions}} \times 1000.
$$

NANN significantly contributes up to 2.4\% CTR and 3.1\% RPM promotion compared with TDM, which demonstrates the effectiveness of our method in both user experience and business benefit.

Moreover, the efficient implementation of NANN introduced in~\Cref{sec:sys} facilitates us to benefit from NANN without sacrificing the RT and QPS of online inference. In production, NANN meets the performance benchmark displayed in~\Cref{tab:perf}. Now, NANN has been fully deployed and provides the online retrieval service entirely in the Taobao display advertising platform.

\section{conclusion}
In recent years, there has been a tendency to tackle the large-scale retrieval problem with deep neural networks. However, these methods usually suffer from the additional training budget and difficulties in using side information from target items because of the learnable index. We propose a lightweight approach to integrating post-training graph-based index with the arbitrarily advanced model. We present both heuristic and learning-based methods to ensure the retrieval quality: 1) our proposed Beam-retrieval can significantly outperform the existing search on graph method under the same amount of computation; 2) we inventively introduce adversarial attack into large-scale retrieval problems to benefit both the retrieval quality and model robustness. Extensive experimental results have already validated the effectiveness of our proposed method. In addition, we summarize in detail the hands-on practices of deploying NANN in Taobao display advertising where NANN has already brought considerable improvements in user experience and commercial revenues. We hope that our work can be broadly applicable to domains beyond recommender system such as web search and content-based image retrieval. In the future, we hope to further uncover the underlying mechanisms that govern the applicability of adversarial attacks to large-scale retrieval problems.

\section{acknowledgements}

We sincerely appreciate Huihui Dong, Zhi Kou, Jingwei Zhuo, Xiang Li and Xiaoqiang Zhu for their assistance with the preliminary research. We thank Yu Zhang, Ziru Xu, Jin Li for their insightful suggestions and discussions. We thank Kaixu Ren, Yuanxing Zhang, Siran Yang, Huimin Yi, Yue Song, Linhao Wang, Bochao Liu, Haiping Huang, Guan Wang, Peng Sun and Di Zhang for implementing the key components of the training and serving infrastructure.

\bibliographystyle{ACM-Reference-Format}
\bibliography{main}

\clearpage
\appendix

\section{Detailed model architecture}

As illustrated in Figure~\ref{fig:network}, our deep neural network mainly consists of four parts: user network, target attention, item network, and score network.
In this section, the detailed model architectures for both the industry dataset and UserBehavior dataset are described respectively.  All features are encoded after the shared embedding layer by mapping from categorical variables to dense vectors. 

\begin{figure}[!t]
    \centering
    \includegraphics[width=0.5\textwidth]{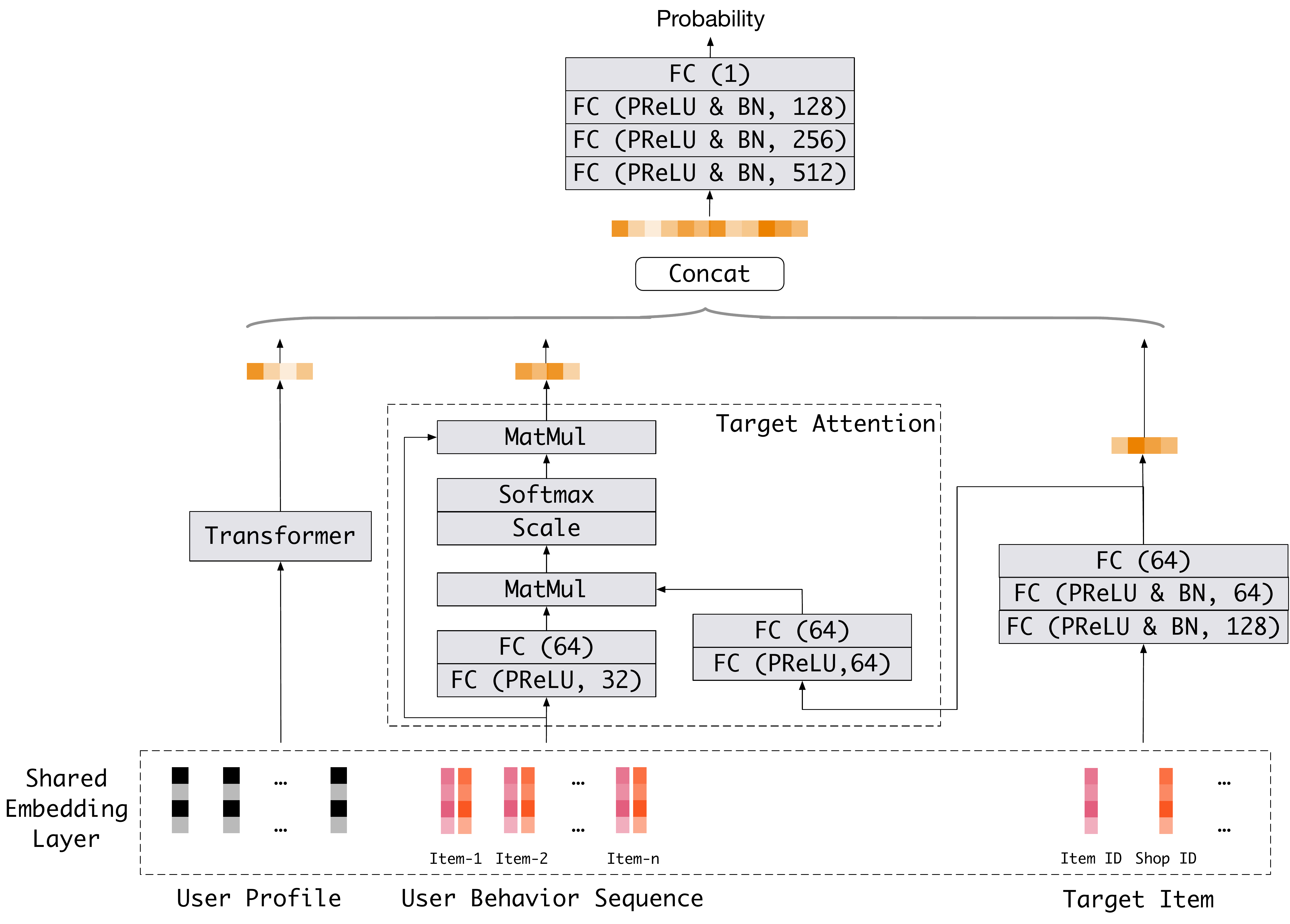}
    \caption{Detailed network architecture for the industrial data of Taobao.}
    \label{fig:detailed_network_industry}
\end{figure}

For the industrial data of Taobao, as shown in ~\Cref{fig:detailed_network_industry}, we build the deep model upon a variety of user and item features, each of which is mapped to an n-dimensional vector of which $n=16$. In the user network branch, the user profile features are fed into the embedding layer to get the user embedding, then a transformer encoder~\footnote{\url{https://www.tensorflow.org/tutorials/text/transformer}} (num\_layers=1, d\_model=16, num\_heads=1, dff=24, seq\_len=51) is leveraged to encode the user's profile with the output shape equal to 16 $\times$ 51. The item network is responsible for modeling the item features and is composed of three fully connected layers (with output shapes equal to 128, 64, 64 respectively), the first two layers are both followed by PRELU and batch normalization. The target attention takes charge of encoding the relevance of the item embedding generated by the item network to the user behavior sequence of which the maximum sequence length is equal to 50. Specifically, we use the scaled dot-product attention to calculate attention scores and output the dense vector by summing over the user behavior sequence with attention scores. The outputs of the user network, target attention, and item network are concatenated and fed into the scoring network to obtain the user-item preference score finally. The score network consists of four fully connected layers (with output shapes equal to 512, 256, 128, 1 respectively), and the first three layers are followed by PRELU and batch normalization.

\begin{figure}[!t]
    \centering
    \includegraphics[width=0.43\textwidth]{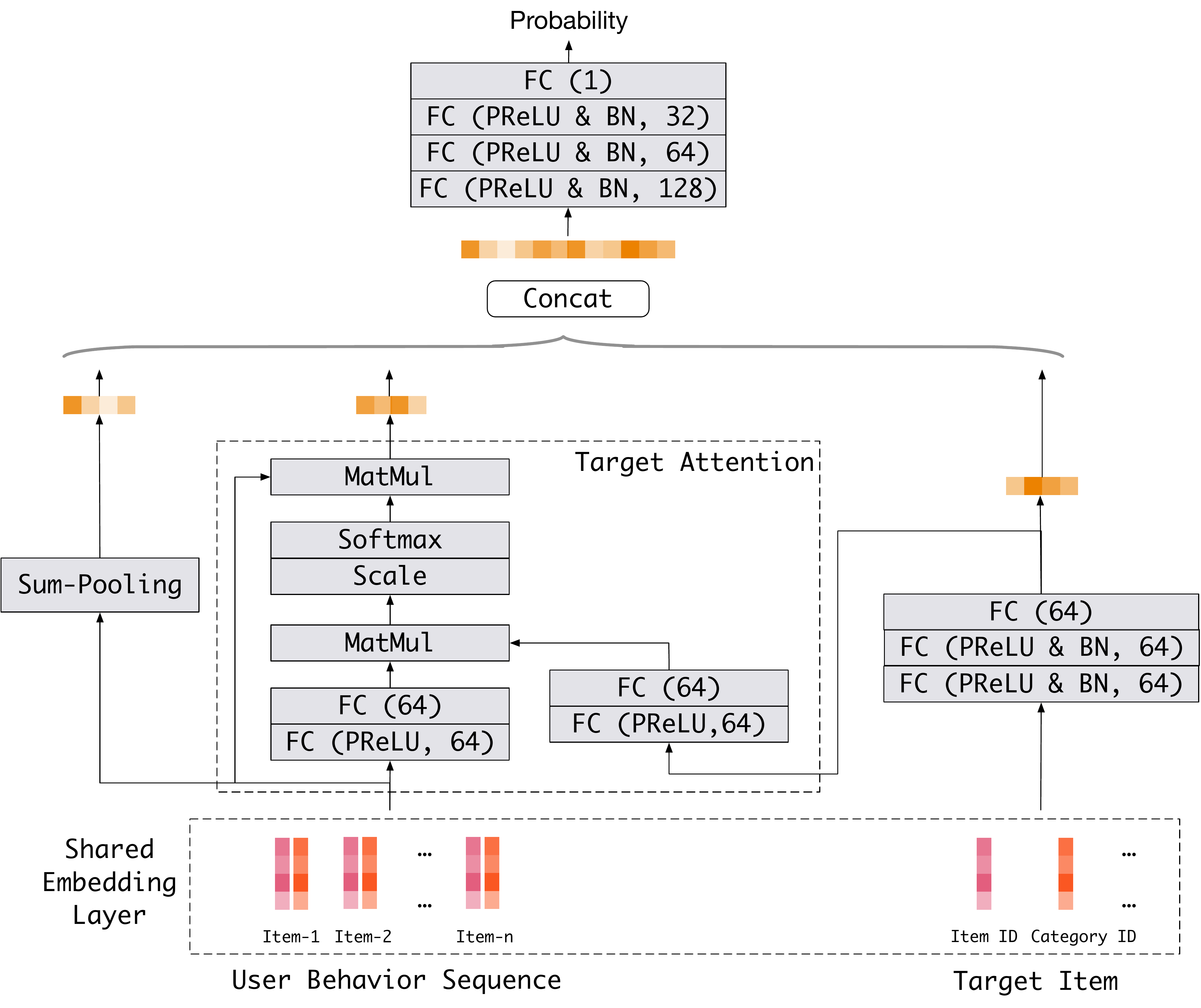}
    \caption{Detailed network architecture for the open-sourced UserBehavior dataset.}
    \label{fig:detailed_network_ub}
\end{figure}

For the UserBehavior dataset, there are no user profile features but the user behavior sequence for each $u \in \mathcal{U}$. And all categorical features are mapped to dense vectors with size=32. As illustrated in~\Cref{fig:detailed_network_ub}, in the user network branch, the user embedding is obtained by sum-pooling over the user behavior sequence. The item network is composed of three fully connected layers (with output shapes equal to 64, 64, 64 respectively) each of which is followed by PRELU and batch normalization. Similar to the industry dataset of Taobao, the scaled dot-product attention is adopted in target attention to calculating attention scores. Then the outputs of these three parts are concatenated and fed into the scoring network to obtain the user-item preference score finally. The score network also consists of four fully connected layers (with output shapes equal to 128, 64, 32, 1 respectively), and the first three layers are followed by PRELU and batch normalization.

\end{document}